\documentclass[aip,jcp,reprint,noshowkeys]{revtex4-1}

\usepackage{amsmath,amssymb,amsfonts,bm,graphicx,dcolumn,xcolor,braket,bigints,physics,siunitx,pgf,upgreek}

\usepackage[utf8]{inputenc}
\usepackage[T1]{fontenc}
\usepackage{libertine,mathpazo}
\usepackage[normalem]{ulem}
\usepackage[version=4]{mhchem}
\usepackage{subcaption,float}
\usepackage{tikz}

\newcolumntype{d}[1]{D{.}{.}{#1}}
\newcommand{\head}[1]{\multicolumn{1}{c}{#1}}
\newcommand\Tstrut{\rule{0pt}{2.5ex}}         
\newcommand\Bstrut{\rule[-1.5ex]{0pt}{0pt}}   

\usepackage{natbib}
\AtBeginDocument{\nocite{achemso-control}}

\usepackage{hyperref}
\hypersetup{colorlinks=true,linkcolor=blue,filecolor=blue,urlcolor=blue,citecolor=blue}

\definecolor{hughgreen}{HTML}{009900}

\definecolor{rhiviolet}{RGB}{148,0,211}

\definecolor{gold}{HTML}{CFB53B}
\tikzset{
    text shadow/.code args={[#1]#2at#3(#4)#5}{
        \pgfkeysalso{/tikz/.cd,#1}%
        \foreach \angle in {0,5,...,359}{
                \node[#1,text=gold] at ([shift={(\angle:.5pt)}] #4){#5};
        }
    }
}
\definecolor{burgundy}{HTML}{902030}

\newcommand{\ETHZ}{Laboratory of Physical Chemistry, Department of Chemistry and Applied Biosciences, ETH Z\"{u}rich, Vladimir-Prelog-Weg 2, 8093 Z\"{u}rich, Switzerland\,}
\newcommand{\UCAM}{Department of Chemistry, University of Cambridge, Lensfield Road, Cambridge, CB2 1EW, U.K.}

\newcommand{\Eh}{\text{E}_{\text{h}}}

\newcommand{\xc}{\hat v_{\text{XC}}}

\newcommand{\br}{\mathbf{r}}

\newcommand{\sig}{\upsigma}
\newcommand{\sigg}{\sig_{\text{g}}}
\newcommand{\sigu}{\sig_{\text{u}}}

\begin{document}

\raggedbottom 

\title{Holomorphic Density-Functional Theory}

\author{Rhiannon A.~Zarotiadis}
\email{rhiannon.zarotiadis@phys.chem.ethz.ch}
\affiliation{\UCAM}
\affiliation{\ETHZ}

\author{Hugh~G.~A.~Burton}
\email{hb407@cam.ac.uk}
\affiliation{\UCAM}

\author{Alex J.~W.~Thom}
\email{ajwt3@cam.ac.uk}
\affiliation{\UCAM}

\date{\today}

\begin{abstract}
Self-consistent-field (SCF) approximations formulated using Hartree--Fock (HF) or Kohn--Sham Density Functional Theory (KS-DFT) both have the potential to yield
multiple solutions.
However, the formal relationship between multiple solutions identified using HF or KS-DFT remains generally unknown.
We investigate the connection between multiple SCF solutions for HF or KS-DFT by introducing a parametrised functional that scales between the two representations.
Using the hydrogen molecule and a model of electron transfer, we continuously map multiple  solutions from the HF potential to a KS-DFT description.
We discover that multiple solutions can coalesce and vanish as the functional changes, forming a direct analogy with the disappearance of real HF solutions along
a change in molecular structure.
To overcome this disappearance of solutions, we develop a complex-analytic extension of DFT --- the ``holomorphic DFT'' approach --- that allows every SCF stationary state to be 
analytically continued across all molecular structures and exchange-correlation functionals.
%
%
\end{abstract}

\maketitle

\section{Introduction}

Solving the electronic Schr\"{o}dinger equation\cite{Schrodinger1926} remains a fundamental challenge in quantum chemistry, in principle enabling an exact theoretical description of chemical properties and reactivity.
However, exact solutions remain elusive beyond the simplest of chemical systems.\cite{Loos2012}
Research has therefore focused on exploiting physical and chemical understanding to develop approximations to the exact electronic structure.\cite{Hirata2012}
At the heart of most approximations lies the self-consistent field (SCF) approach,\cite{SzaboBook} usually through the form of Hartree--Fock (HF) or Kohn--Sham Density-Functional Theory (KS-DFT).
However, beyond simply providing a `reference state' for correlated approaches,\cite{HelgakerBook} the SCF approximation is itself a rich theory with the potential to provide chemical insights into excited states\cite{Gilbert2008} and reactive bond-breaking processes.\cite{Jensen2018}

SCF methods are usually presented as iterative approaches. 
On each iteration, the electron density obtained from the previous step is used to build an approximate electronic potential that is then used to re-optimise the electron density or wave function as an input for the next iteration. 
This process is repeated until self-consistency is reached.\cite{SzaboBook}
Alternatively, the SCF energy can be considered as a non-linear function of the one-electron density, with the global minimum corresponding to the approximate electronic ground state.
Besides the global minimum, this non-linear function can possess several stationary points that each represent an optimal SCF state and correspond to local minima, maxima, or saddle points of the SCF energy.\cite{Fukutome1971,Stanton1968,Thom2008}

Historically, the existence of multiple SCF solutions has been considered as a computational obstacle, particularly when the lowest energy HF ground state is required\cite{Thouless1960,Adams1962,Cizek1967,Seeger1977} or during \textit{ab initio} molecular dynamics simulations involving molecules with multiple low-lying states.\cite{Vaucher2017}
Alternatively, recent research has developed and exploited physical interpretations of multiple SCF solutions themselves. \cite{Gilbert2008, Barca2014}
For example, encouraged by new computational methods that make identifying higher energy stationary points relatively routine,\cite{Gilbert2008,Thom2008} multiple SCF solutions have been used as mean-field approximations to excited states.\cite{Gilbert2008,Besley2009,Barca2014}
Furthermore, the similarities between dominant electron configurations in strongly correlated molecules and multiple HF states have motivated their use as a basis for multireference ground- and excited-state wave functions.
Since each HF solution comprises an independent set of molecular orbitals (MOs), these multireference calculations take the form of a nonorthogonal configuration interaction (NOCI).\cite{Thom2008,Sundstrom2014,Burton2019c}
However, in many cases, these applications have been hindered by the disappearance of SCF solutions at so-called Coulson--Fischer points as molecular structure changes, with the low-lying unrestricted (UHF) states in \ce{H2} providing the archetypal example.\cite{Coulson1949}

Recently, holomorphic Hartree--Fock (h-HF) theory has been  developed as a method for extending real HF states into the complex plane, beyond the Coulson--Fischer points at which they vanish in conventional HF.\cite{Hiscock2014, Burton2016, Burton2018, Burton2019a}
In h-HF, the complex-conjugation of orbital coefficients is removed from the conventional energy function to define a complex-analytic analytic continuation of the conventional HF equations.\cite{Hiscock2014, Burton2016, Burton2018}
The resulting h-HF stationary points then exist across all molecular geometries, allowing methods such as NOCI to be generalised as alternatives to conventional multireference approaches such as the complete active-space SCF framework.\cite{Burton2016,Burton2019c}
Furthermore, h-HF theory has provided extensive insight into the fundamental nature of multiple SCF solutions, revealing that discrete HF solutions can be connected as one continuous structure in the complex plane\cite{Burton2019a} and allowing new symmetries to be identified that ensure real HF energies with non-Hermitian Fock matrices.\cite{Burton2019b}

HF theory represents only one example of the SCF approximation and, as a mean-field method,  fails to accurately reproduce the electron-electron correlation that is essential for the correct prediction of chemistry.\cite{SzaboBook,electronCorrelation2, tew, electronCorrelation}
An alternative approach, DFT has been developed to capture electron correlation in the SCF framework.\cite{HK, KS}
DFT approximations generally utilise empirical energy functionals of the electron density to describe the most physically relevant electron correlation effects. \cite{KohnNobel, DFTperspective}
The relative accuracy, low-order scaling, and computational simplicity of DFT has led to its widespread application as one of the most popular electronic structure techniques.\cite{100papers}

In principle, the SCF nature of DFT can also produce multiple stationary points with the same potential applications as multiple HF states.
For example, higher energy solutions can be exploited as approximations to excited states through the $\Delta$SCF framework.\cite{Theophilou1979,Gunnarsson1976,Gorling1999,Jones1989}
However, the behaviour of multiple DFT solutions as the molecular structure or chosen functional changes, and their relationship to standard HF solutions, appears relatively unexplored.
This lack of knowledge is both surprising and concerning given that certain DFT solutions are known to also disappear at Coulson--Fischer points, leading to kinks and discontinuities along the corresponding potential energy surfaces.\cite{MoriSanchez2014} 
We therefore believe that a detailed investigation into the relationship between multiple HF and DFT solutions is well overdue.

In this work, we aim to extend our understanding of multiple DFT solutions by following solutions along a path between HF theory and DFT.
In Section \ref{sec:motivation} the relationship between the solutions in HF and two fundamental DFT functionals are investigated for a typical electron transfer model.\cite{Jensen2018}
We find that DFT solutions can coalesce and vanish in exactly the same manner as real HF solutions.
Motivated by this discovery, in Section \ref{sec:theory} and beyond we investigate a holomorphic extension of DFT with the potential to analytically continue DFT solutions across all molecular structures.
In doing so, we reveal fundamental relationships between the SCF states of DFT functionals and those of HF, laying the foundation for a more informed exploitation of multiple DFT solutions in chemical applications.


\section{Scaling Between Hartree--Fock and DFT}
\label{sec:motivation}

HF theory provides the foundation for almost all sophisticated wavefunction-based electronic structure calculations.
The inadequacies in the HF description of molecules have  been well-investigated and thus, although it does not produce the exact electronic energy, crucial understanding can be obtained from a HF calculation.\cite{SzaboBook}
It is therefore interesting to investigate how SCF states evolve from this approximate but well-defined HF description to (hopefully) more accurate, but often empirical DFT functional.

\begin{figure}[b!]
	\center
	\includegraphics[width=\columnwidth]{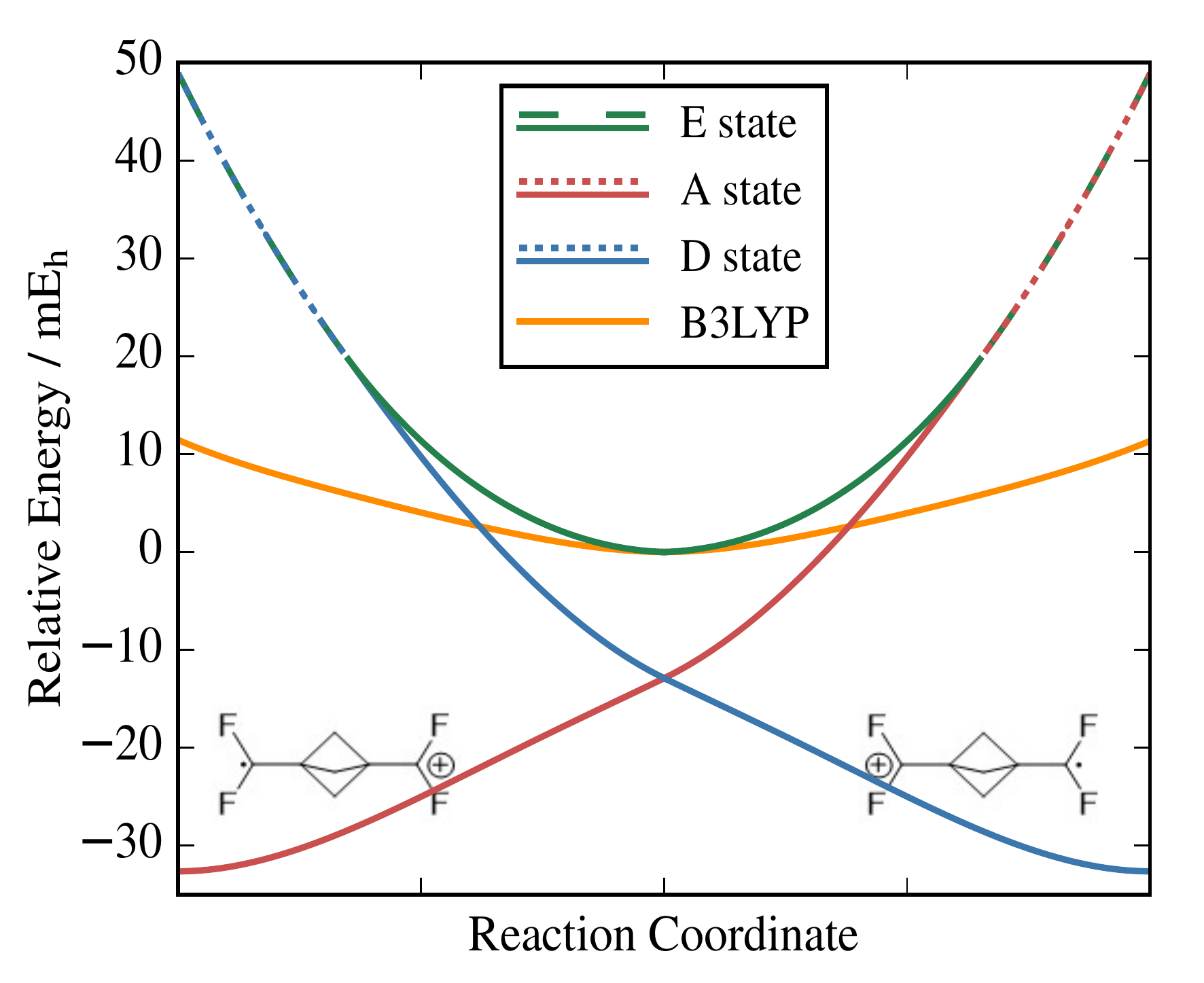}
	\caption[Electronic energies of the electron transfer model]
	{   \label{fig:reactionEnergy}
		Electronic energies along the electron transfer reaction trajectory for the model electron transfer system shown.
		The donor (D) and acceptor (A) states are interconverted by symmetry at the transition state, while the E state is symmetric across all geometries.
		Real HF energies (solid lines) have previously been reported in Ref.~\onlinecite{Jensen2018} and are plotted relative to the minimum energy of the E state.
		Dashed lines indicate the holomorphic continuation of a given state into the complex plane, where only the real component of the h-HF energy is plotted.
		Only one low-energy state can be identified using B3LYP-DFT, and this is plotted relative to its minimum energy.
	}
\end{figure}

The electron transfer model \ce{C_7H_6F_4^{.+}} studied by Jensen \textit{et al.}\cite{Jensen2018}\ provides an interesting case-study for comparing the HF and DFT approximations.
In this model, a single electron transfers from one carbon-di-fluoride group to its symmetric counterpart along a collective reaction coordinate (see Figure~\ref{fig:reactionEnergy}).
When applying HF theory, three chemically relevant SCF states can be identified corresponding to the symmetry-broken diabatic electron donor (D) and acceptor (A) configurations, and a third delocalised symmetric state (E) that represents the transferring electron.
All three states are stationary solutions to the real HF equations at the minimum energy crossing point (MECP) of the D and A states.
However, as the molecule distorts away from the MECP towards the donor or acceptor structure, the A/D and E configurations coalesce and vanish.
These properties of the real HF states were described in detail in Ref.~\onlinecite{Jensen2018}, although we can now report the existence of the complex-valued h-HF extensions shown in Figure~\ref{fig:reactionEnergy}.

Alternatively, the B3LYP-DFT functional only yields one low-lying stationary state with an electron density at the MECP that most closely resembles the E state (orange line in Fig.~\ref{fig:reactionEnergy}).
This DFT solution predicts a single energy minimum along the reaction coordinate, providing a contrasting picture to the electron transfer predicted by the symmetry-broken HF states (although the reaction trajectory is not optimised for the B3LYP energy).
It is not immediately obvious which  features of the HF and B3LYP potentials cause these qualitatively different energy surfaces, or which potential would provide the most faithful representation of the electron transfer process.
However, it is surprising that the multiple symmetry-broken HF states, which appear to resemble diabatic electron transfer configurations, appear completely absent in the B3LYP-DFT description.

To understand \emph{why} these additional states no longer exist using B3LYP-DFT,  we follow the real HF states as the SCF approximation is continuously scaled from  HF to DFT.
Unless otherwise stated, all further calculations are performed at the MECP geometry where the relevant HF states exist and the D and A states become degenerate.
For simplicity, we consider the minimal STO-3G basis rather than the cc-pVDZ basis used by Jensen \textit{et al.},\cite{Jensen2018} although this does not change any qualitative features of the SCF solutions.

The relationship between HF theory and DFT can be seen when the HF\cite{SzaboBook} and KS-DFT\cite{KS, KohnNobel} equations are written respectively as
\begin{subequations}
\begin{align}
\left[-\frac{1}{2} \nabla^2 + v_\text{eN}(\mathbf{r}) + j(\mathbf{r}) + \hat k(\mathbf{r})\right] \phi_i^\text{HF}(\mathbf{r}) 
&= \epsilon_i^\text{HF} \phi_i^\text{HF}(\mathbf{r}) 
\label{HFscf}
\\
\left[-\frac{1}{2} \nabla^2 + v_\text{eN}(\mathbf{r}) + j(\mathbf{r}) + v_\text{XC}(\mathbf{r})\right]\phi_i^\text{KS}(\mathbf{r}) 
&= \epsilon_i^\text{KS} \phi_i^\text{KS}(\mathbf{r}). 
\label{KSscf}
\end{align} 
\end{subequations}
Here, the electronic kinetic operator $-\frac{1}{2}\nabla^2$, electron-nuclear potential $v_\text{eN}(\mathbf{r})$, Coulomb potential $j(\mathbf{r})$ and the respective exchange operator $\hat k(\mathbf{r})$ and exchange correlation potential $v_\text{XC}(\mathbf{r})$ are applied to the HF orbitals $\phi_i^\text{HF}(\mathbf{r})$ or the KS orbitals $\phi_i^\text{KS}(\mathbf{r})$ to obtain the Lagrange multipliers $\epsilon_i^\text{HF}$ or $\epsilon_i^\text{KS}$.
Each molecular orbital is expanded in terms of the $m$-dimensional finite basis set with the orbital coefficients $c_{\cdot i}^{\mu \cdot}$ as 
	\begin{equation}
		\phi_i(\mathbf{r}) = \sum_{\mu}^{m} \chi_\mu (\mathbf{r}) c_{\cdot i}^{\mu \cdot},
	\end{equation}
where the atomic orbitals (AOs) and MOs are given by $\chi_\mu (\mathbf{r})$ and $\phi_i(\mathbf{r})$ respectively.
Herer we employ the nonorthogonal tensor notation defined Ref.~\onlinecite{HeadGordon1998}, and apply the Einstein summation convention whenever summation is not indicated explicitly.
These HF and DFT equations can be conceptually unified by introducing a parametrised exchange-correlation operator $\xc(\br; q)$ that interpolates between HF exchange and DFT exchange-correlation with the form
\begin{equation}
\xc(\br; q) = (1-q)\, \hat k(\br) + q\, \nu_{\text{XC}}(\br).
\label{convScaling}
\end{equation}
A scaling parameter of $q=0$ corresponds to a pure HF calculation, while $q=1$ refers to a pure DFT calculation.
Individual ground- and excited-state SCF solutions can then be traced between different functionals using the maximum overlap method (MOM),\cite{Gilbert2008, Barca2018} where non-\textit{Aufbau} optimisation is achieved by selecting the new occupied orbitals on each SCF iteration according to their overlap with the occupied orbitals on the previous iteration.

As the physical functional evolves, the relationship between different SCF solutions can be visualised by considering a similarity measure for two SCF states $\kappa$ and $\lambda$.
Here we apply the distance measure introduced by Thom \textit{et al.},\cite{Thom2008} which uses the density matrices ${^{\kappa}\kern-0.15em P}$ and ${^{\lambda}\kern-0.15em P}$ to define the distance between two $N$-electron states as
\begin{equation}
d_{\kappa \lambda}^2 
= \norm{ {^{\kappa}\kern-0.15em P} - {^{\lambda}\kern-0.15em P} }^2  
= N  -  {^{\kappa}\kern-0.15em P}^{\mu \nu} S_{\nu \sigma} {^{\lambda}\kern-0.15em P}^{\sigma \tau} S_{\tau \mu}.
\label{eq:sqDistance}
\end{equation}
The density matrix for a given state $\kappa$ is defined in terms of the occupied MO coefficients $({^{\kappa}\kern-0.15em c})_{\cdot i}^{\mu \cdot}$ as 
\begin{equation}
{^{\kappa}\kern-0.15em P}^{\mu \nu} 
= ({^{\kappa}\kern-0.15em c}^{\vphantom{*}})_{\cdot i}^{\mu \cdot} ({^{\kappa}\kern-0.15em c}^{*})_{i \cdot}^{\cdot \nu},
\end{equation}
where $S_{\nu \sigma}$ denotes the AO overlap matrix, and the summation of repeated indices is implicit.
The second equality in Eq.~\eqref{eq:sqDistance} bounds the distance measure as $d_{\kappa \lambda}^2 \in [0,N]$, giving the distance measure in units of `electron number'.

\begin{figure*}
    \centering
	\begin{subfigure}[b]{0.45\textwidth}
		\includegraphics[width=\columnwidth,trim=70pt 70pt 80pt 60pt, clip]{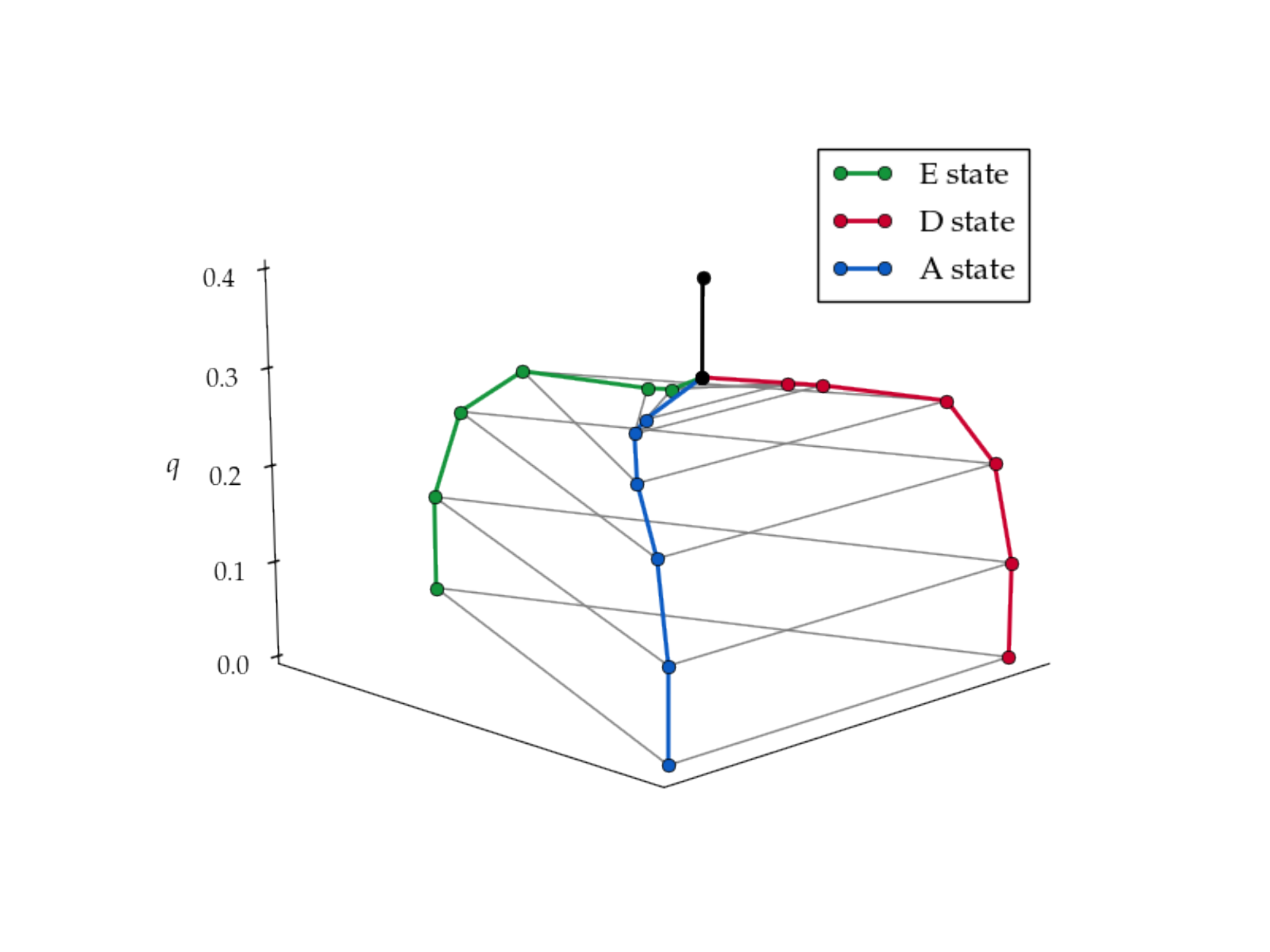}
		\subcaption{}
		\label{subfig:LDA_mecp}
	\end{subfigure}
	\begin{subfigure}[b]{0.45\textwidth}
		\includegraphics[width=\columnwidth,trim=70pt 70pt 80pt 60pt, clip]{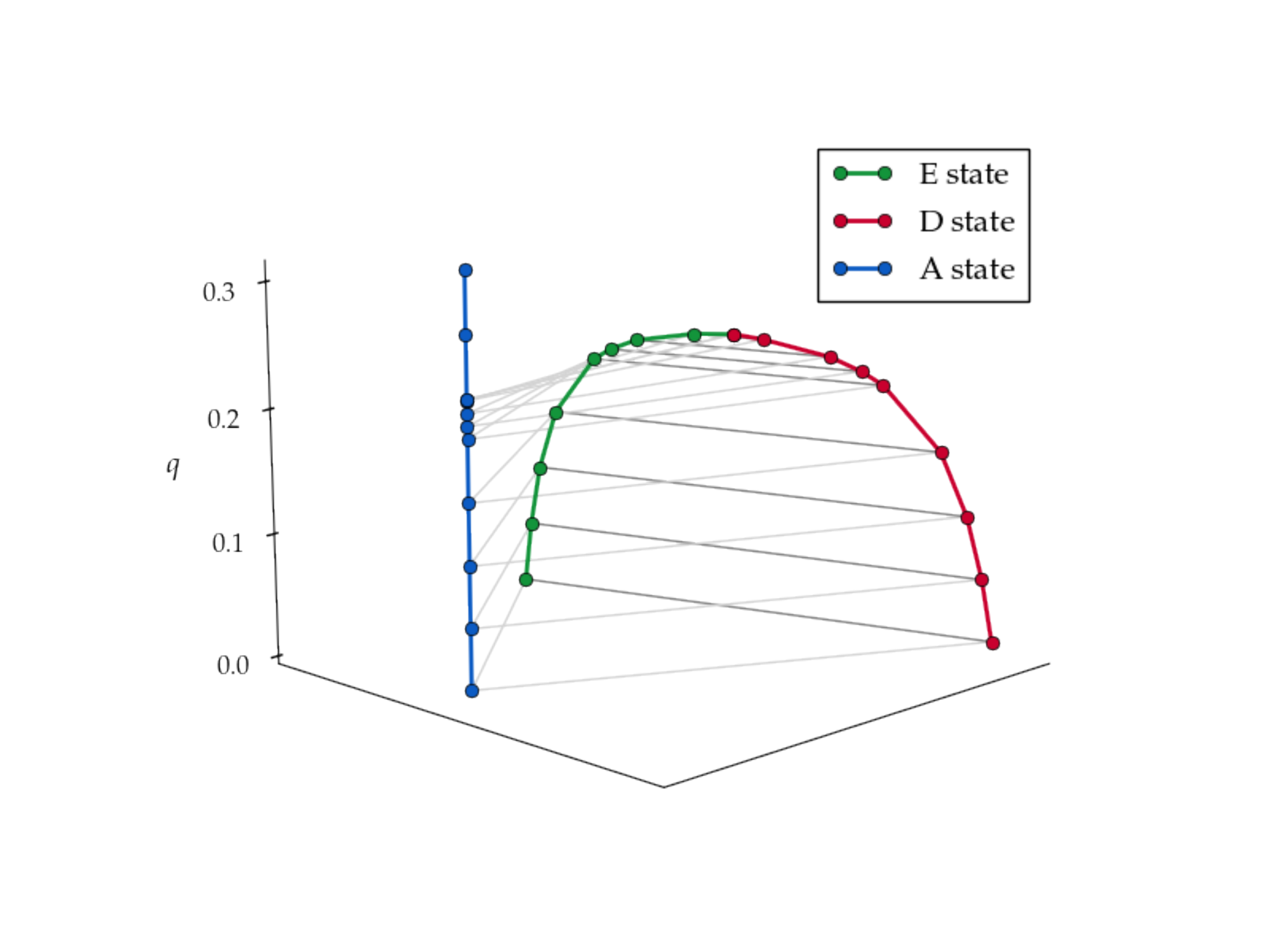}
		\subcaption{}
		\label{subfig:LDA_not_mecp}
	\end{subfigure}
	\caption{Relative distances of the SCF solutions as the exchange correlation functional is scaled from exact HF exchange $k$ to the LDA-exchange functional.\cite{bookLDA}.  
		The grey lines between solutions at each plane of constant $q$ correspond to the square-root of the inter-state distances~\eqref{eq:sqDistance}.
		(\subref{subfig:LDA_mecp}) At the MECP, the E (green line), D (red line), and A (blue line) states all simultaneously coalesce at approximately $q=0.3$ to leave a single SCF solution (black line).
		(\subref{subfig:LDA_not_mecp}) When the structure is distorted towards the acceptor structure, the E (green line) and D (red line) states coalesce and both vanish at approximately $q=0.3$, while the A state (blue line) remains independent across all values of $q$.
	}
	\label{xOnlyLDADist}
\end{figure*}

We first consider scaling between HF and the analytic Local Density Approximation (LDA) exchange functional.\cite{bookLDA}
At the MECP, the three  SCF states A, D, and E simultaneously coalesce as $q$ scales between the HF and LDA-exchange description, as demonstrated in Figure~\ref{subfig:LDA_mecp} where the distance measure between the states falls to zero at the point of coalescence.
This three-fold coalescence occurs at a ``confluence'' point,\cite{Fukutome1975,Burton2018} where the degenerate A and D solutions coalesce with the higher energy E state to leave only the E state for larger values of $q$ (black line).
In contrast, when the molecular geometry is marginally distorted towards the acceptor structure, the symmetry-broken D state and the symmetric E state coalesce and vanish at a ``pair annhilation'' point,\cite{Fukutome1975,Burton2018} while the A state can be traced continuously from HF to LDA (Figure~\ref{subfig:LDA_not_mecp}).
This observation indicates that the single LDA solution is not a direct mirror of the E state in HF, but evolves continuously from the A state to the D state as the molecular structure changes.
The LDA-DFT stationary state therefore appears to behave as an adiabatic state (as one would expect for states obtained using the exact functional), in contrast to the diabatic behaviour of the multiple HF states.

But why do DFT functionals yield adiabatic states rather than the multiple symmetry-broken diabatic states observed using HF?
One possible explanation for the coalescence of SCF states is the self-interaction error (SIE), which is a well-known problem of not only LDA-DFT but also more elaborate DFT functionals such as B3LYP.\cite{Perdew1981, Zhang1998, Lundberg2005, MardirossianReview2017}
It has been shown that the exchange contribution of different DFT functionals may include dynamic correlation effects through the SIE, and these effects can dominate the change of electron density between different correlation functionals.\cite{Graefenstein2009}

\begin{figure*}[thb!]
	\begin{subfigure}[b]{0.32\textwidth}
		\includegraphics[width=\textwidth,trim=70pt 70pt 80pt 40pt, clip]{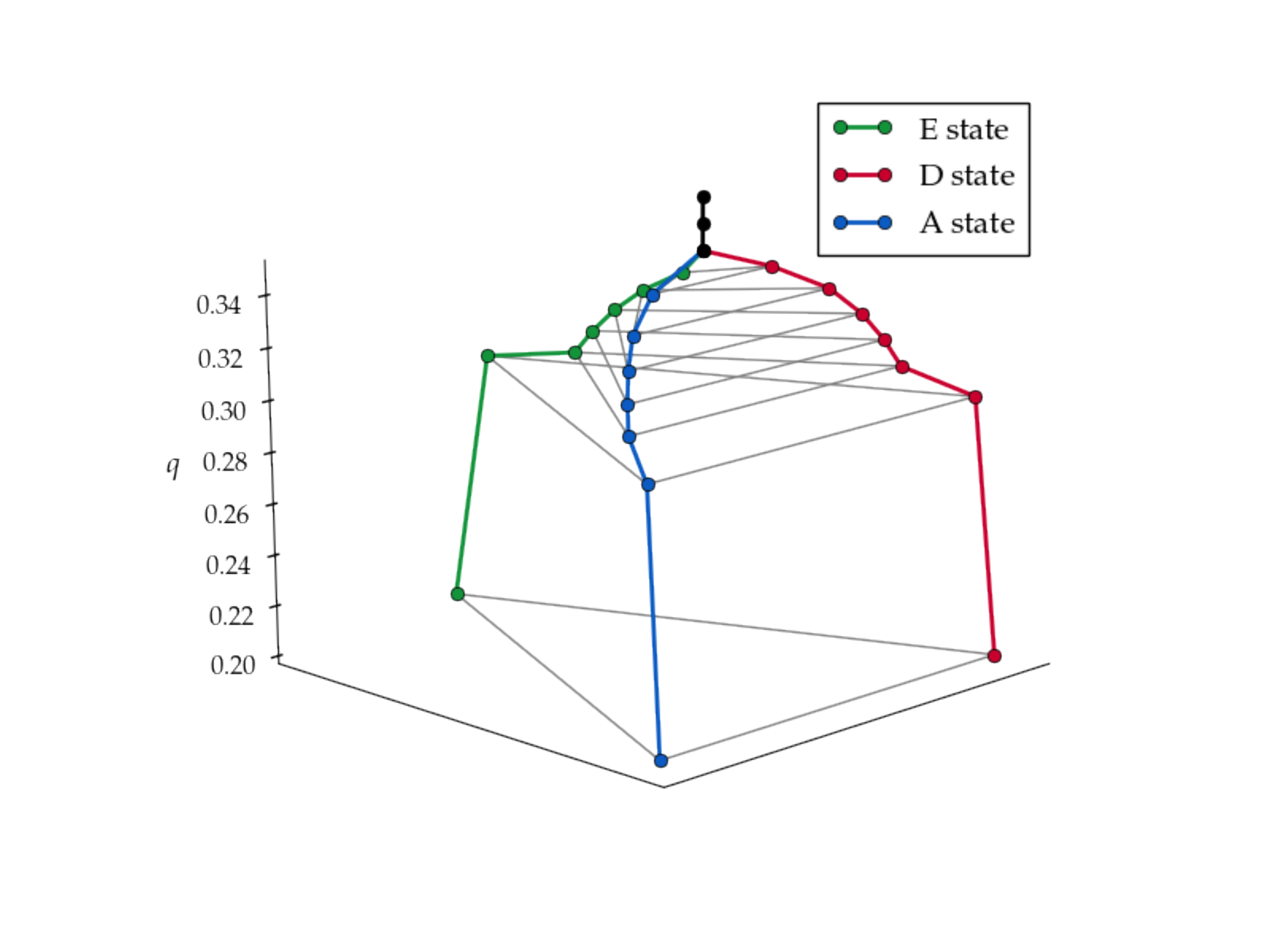}
		\subcaption{}
		\label{xcOnlyDist}
	\end{subfigure}
	\begin{subfigure}[b]{0.32\textwidth}
		\includegraphics[width=\textwidth,trim=70pt 70pt 80pt 40pt, clip]{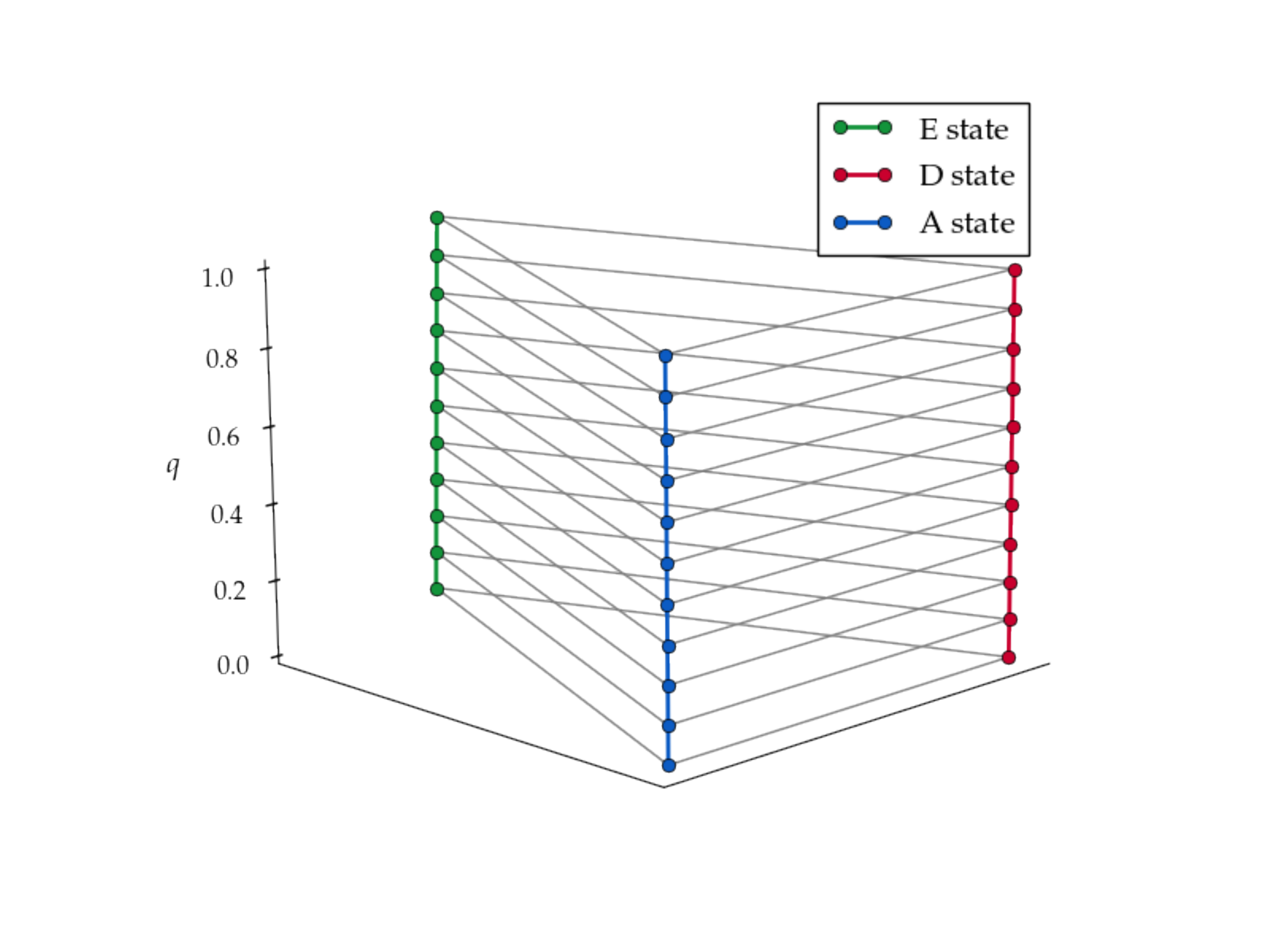}
		\subcaption{}
		\label{cOnlyDist}
	\end{subfigure}
	\begin{subfigure}[b]{0.32\textwidth}
		\includegraphics[width=\textwidth,trim=70pt 70pt 80pt 40pt, clip]{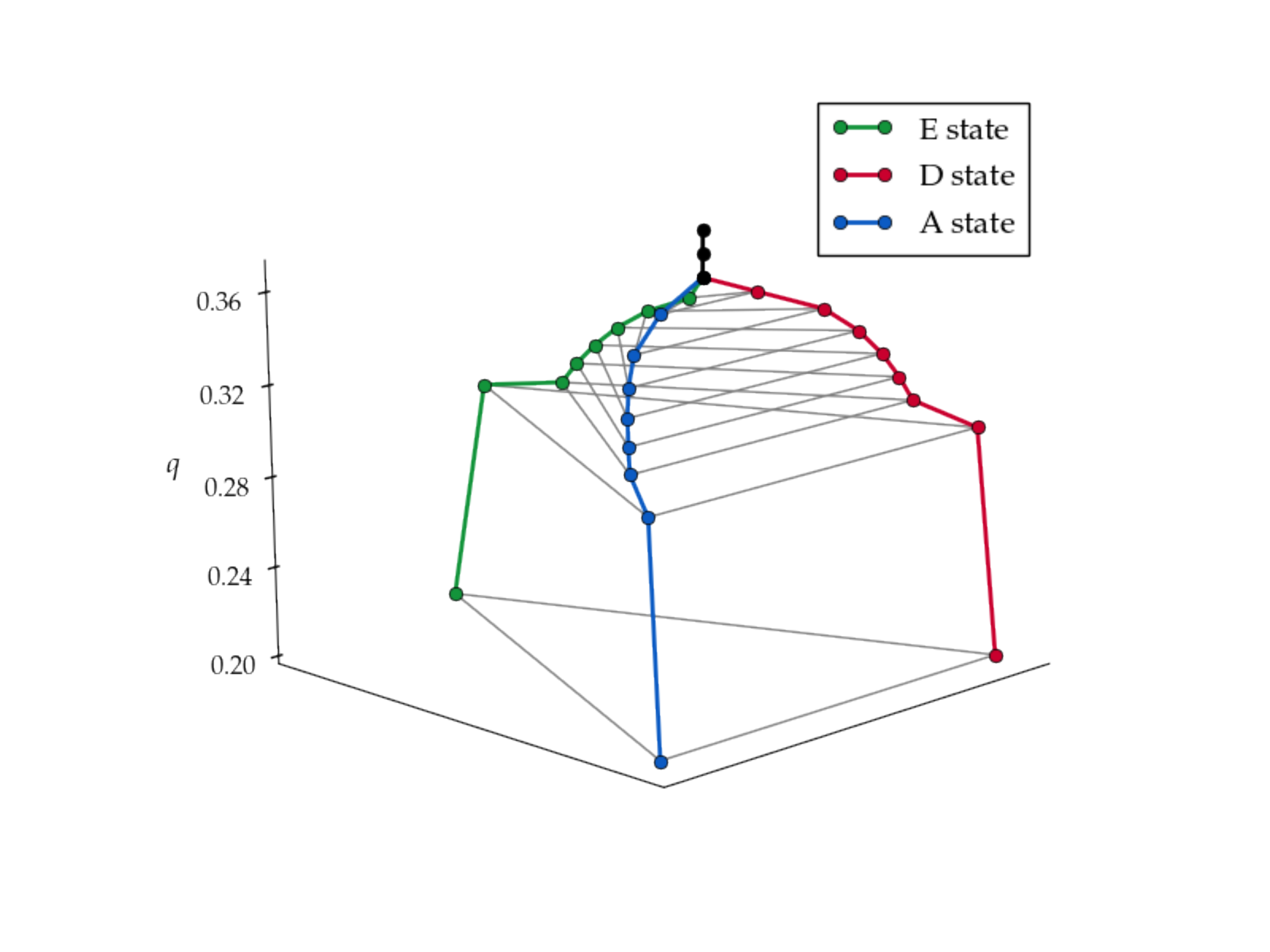}
		\subcaption{}
		\label{xOnlyDist}
	\end{subfigure}
	\caption{
		Relative distances for the three SCF states in the different scaling modes.
		(a) Full B3LYP exchange-correlation functional.\cite{B88, LYP}
		(b) LYP correlation functional:\cite{LYP} No coalescence of the SCF states is observed and the distance measure remains virtually unchanged.
		(c) B3LYP exchange functional:\cite{B88} Coalescence of the three SCF states occurs in a similar manner to the introduction of the full B3LYP functional.
	}
\end{figure*}

To understand how different components in a DFT functional affect the existence of multiple SCF solutions, we scale the same electron transfer model between HF and the popular B3LYP functional\cite{B88, LYP, 100papers} and find that it leads to the same pattern of coalescence between the three SCF states.
We then decompose the B3LYP-DFT functional into its constituent exchange and correlation energy contributions and consider scaling between the HF and B3LYP-DFT potential using only the LYP correlation term (Figure~\ref{cOnlyDist}) or the exchange description (Figure~\ref{xOnlyDist}), and the scaling between HF and the full B3LYP functional (Figure~\ref{xcOnlyDist}).

When only the LYP correlation term is included (Figure~\ref{cOnlyDist}), the SCF states remain distinct for all values of the scaling parameter and no coalescence is observed.
In contrast, introducing only the exchange contribution (Figure~\ref{xOnlyDist}) causes all three SCF states to coalesce at approximately the same scaling level as the full B3LYP picture.
This coalescence demonstrates that the exchange correlation functional provides the driving force for the coalescence of states as the SCF approximation is scaled between HF and B3LYP.
Furthermore, inspecting the individual components of the energy (not shown) reveals that the magnitude of the total exchange contribution decreases as one moves from HF to B3LYP.
The coalescence of the symmetry-broken SCF states is therefore driven by the overall strength of the exchange interaction.
This result is entirely consistent with other instances of symmetry-breaking in HF theory, for example the emergence of spin-density waves in antiferromagnetic materials.\cite{Slater1951}

The electron transfer model reveals that smoothly changing the exchange-correlation functional can lead to the coalescence of SCF solutions in exactly the same way as changing the molecular structure.
Furthermore, we have found that the strength of the exchange interaction is a key factor in controlling whether several symmetry-broken SCF stationary states can be identified.
While there are three distinct solutions using the HF exchange functional, a small perturbation of this exchange description towards DFT is sufficient to collapse these diabatic solutions onto one adiabatic state. 
However, to completely connect the SCF states from HF to DFT, we require an approach that extends SCF solutions beyond the scaling levels at which SCF states vanish.
Following the framework of holomorphic HF, we believe that a holomorphic extension to DFT will allow SCF states to be analytically continued into the complex plane, 
and developing such a method forms the focus for the remainder of this paper.

\section{Holomorphic Density Functional Theory}
\label{sec:theory}

Before deriving a holomorphic extension to DFT, we  first review the h-HF approach itself.\cite{Hiscock2014,Burton2016,Burton2018,BurtonThesis}
The original motivation  for h-HF theory is rooted in the desire to extend real HF states across all molecular geometries and construct a continuous basis for multireference NOCI calculations.\cite{Thom2009}
To extend real HF states beyond the Coulson--Fischer points at which they vanish, the h-HF energy function $\tilde{E}$ is formulated as a complex-analytic extension of the real HF energy,\cite{Hiscock2014,Burton2016} given for a closed-shell $N$-electron system with $m$ basis functions as
\begin{align}
\tilde{E}
&= E_{\text{nuc}}  + \sum_{\mu \nu}^m \tilde{P}^{\nu \mu}  \qty( 2h_{\mu \nu} + 2 j_{\mu \nu} - k_{\mu \nu} ),
\label{eq:HoloE}
\end{align}
where the holomorphic density matrix has been introduced as
\begin{equation}
\tilde{P}^{\nu \mu} = \sum_{i}^{N/2} c^{\nu \cdot}_{\cdot i} c^{\cdot \mu}_{i \cdot}.
\label{eq:HoloDensity}
\end{equation}
Here, $h_{\mu \nu}$ denotes the one-electron integrals and the self-consistent Coulomb and exchange matrices are defined in terms of two-electron integrals as
\begin{subequations}
\begin{align}
	j_{\mu \nu} = \sum_{\sigma \tau}^{m}  \langle \mu \sigma |\nu \tau\rangle \tilde{P}^{\tau \sigma},
	\\
	k_{\mu \nu} = \sum_{\sigma \tau}^{m}  \langle\mu \sigma | \tau \nu\rangle \tilde{P}^{\tau \sigma}.	
\end{align}
\end{subequations}
By removing the complex-conjugation of orbital coefficients in the holomorphic density matrix \eqref{eq:HoloDensity}, the energy function \eqref{eq:HoloE} satisfies the Cauchy--Riemann conditions\cite{FischerBook} and becomes a complex-analytic polynomial of the orbital coefficients, in contrast to the standard formulation of the HF energy using complex molecular coefficients (see Ref.~\onlinecite{Pople1971}). 
As a result, every real HF state remains a stationary point of the h-HF energy and, when a real HF state disappears, its holomorphic counterpart continues to exist with complex-valued orbital coefficients.\cite{Hiscock2014, Burton2016, Burton2018,Burton2019c}
Crucially, the fact that $\tilde{E}$ is a polynomial of only the orbital coefficients and not their complex conjugates is sufficient to allow a rigorous proof that the number of h-HF states for two-electron systems is constant for all molecular structures.\cite{Burton2018,BurtonThesis}

As a complex-analytic extension to the real HF energy, the operator form of the holomorphic energy function remains the same as the conventional HF energy function~\eqref{HFscf}, and can still be written in terms of the one-electron integrals $h_{\mu \nu}$, and the self-consistent integrals $j_{\mu \nu}$ and $k_{\mu \nu}$ representing the Coulomb interaction and the exact exchange term respectively.
Furthermore, the holomorphic electron density matrix $\tilde{P}^{\nu \mu}$ can be considered as a complex-analytic extension of the real density matrix, although the holomorphic density matrix is complex-symmetric rather than Hermitian.\cite{Burton2018}

In analogy to h-HF theory, we expect that the holomorphic DFT (h-DFT) energy in the Kohn--Sham formalism should also form a complex-analytic function of only the orbital coefficients (and not their complex conjugates) to ensure that its stationary states never disappear.
However, the form of the DFT exchange-correlation functional is not known \textit{a priori} and the exchange-correlation functional is not necessarily a pure polynomial of the MO coefficients. 
Instead, we retain the DFT tradition of focussing on the electron density and define the holomorphic electron density $\tilde{\rho}\qty(\br)$. 
We require this holomorphic electron density to depend on only the orbital coefficients $\{c_{\cdot i}^{\mu \cdot}\}$ and not their complex-conjugates by defining $\tilde{\rho}\qty(\br)$ as
\begin{equation}
\begin{split}
\tilde{\rho}(\br) 
&= \sum_i^{N/2}  \bigg(\sum_{\mu}^{m} \chi_{\mu}\qty(\br) c^{\mu \cdot}_{\cdot i} \bigg) \bigg(\sum_{\nu}^{m} c^{\cdot \nu}_{i \cdot} \chi_{\nu}\qty(\br) \bigg)
\\
&= \sum_{\mu \nu}^{m}  \tilde{P}^{\mu \nu} \chi_{\mu}\qty(\br) \chi_{\nu}\qty(\br),
\end{split}
\label{holoDensity}
\end{equation}
where the holomorphic density matrix, $\tilde{P}^{\mu \nu}$, is given by Eq.~\eqref{eq:HoloDensity}.

\subsection{Holomorphic Density Fitting}
\label{sec:DensityFitting}

Following the initial h-HF investigation,\cite{Hiscock2014} we first attempt to identify h-DFT solutions by analytically solving the h-DFT equations.
Since exchange-correlation functionals used in DFT often contain fractional exponents of the electron density, we retain a complex-analytic polynomial 
form by introducing density fitting methods.\cite{densFit1, densFit2, densFit3, Dunlap1979, densFit5, densFit6, densFit7, densFit8, densFit9, densFit10, densFit11, densFit12}
For the specific case of the LDA exchange energy functional $E_\text{X}^{\text{LDA}}$, given by
\begin{equation}
E_{\text{X}}^{\text{LDA}}[\rho] =-\frac{3}{4} \left( \frac{3}{\pi} \right)^{\frac13} \int \rho(\mathbf{r})^{\frac43} \text{ d$^3$\textbf{r}},
\end{equation}
we express the holomorphic cubed-root electron density $\tilde{\rho}\qty(\br)^{\frac13}$ in a polynomial form as
\begin{equation}
\tilde{\rho}\qty(\br)^{\frac13} = \sum_{\alpha}^{\infty} f^{\alpha} \xi_{\alpha}(\mathbf{r})
\label{eq:CubedRootDen}
\end{equation}
where the density-fitting basis functions, $\xi_{\alpha}(\mathbf{r})$, are different to the AO basis.
The holomorphic LDA exchange functional $\tilde{E}_\text{X}^\text{LDA}$ is then expressed as
\begin{align}
\hspace{-2em}&\tilde{E}_\text{X}^\text{LDA}[\tilde{\rho}^{\frac13}] 
= - \frac{3}{4} \left( \frac{3}{\pi} \right)^{\frac13} \hspace{-0.5em}\bigintsss \left( \sum_{\alpha} f^{\alpha} \xi_{\alpha} (\br) \right)^4 \text{d}^3\br
\\
&= - \frac{3}{4} \left( \frac{3}{\pi} \right)^{\frac13} \hspace{-0.5em} \bigintsss \sum_{\nu \mu \sigma \tau} f^{\alpha} f^{\beta} f^{\gamma} f^{\delta} \xi_{\alpha}(\br) \xi_{\beta}(\br) \xi_{\gamma}(\br) \xi_{\delta}(\br) \text{d}^3\br \nonumber
\end{align}
leading to a fourth-order polynomial in the new coefficient set $\{f^{\alpha}\}$.
The expansion \eqref{eq:CubedRootDen} now allows the holomorphic DFT energy $\tilde{E}$ to be expressed as a complex-analytic polynomial of the MO coefficients $\{c_{\cdot i}^{\mu \cdot}\}$ and the cubed-root electron-density expansion coefficients $\{f^{\alpha}\}$.
We therefore expect that this polynomial energy functional will enable complex-analytic continuations of SCF states in a combined HF and DFT framework to be identified beyond the points where real SCF states coalesce and disappear, and our current implementation is described in Appendix~\ref{sec:appendix}.

\begin{figure}[hbt!]
\includegraphics[width=0.96\columnwidth]{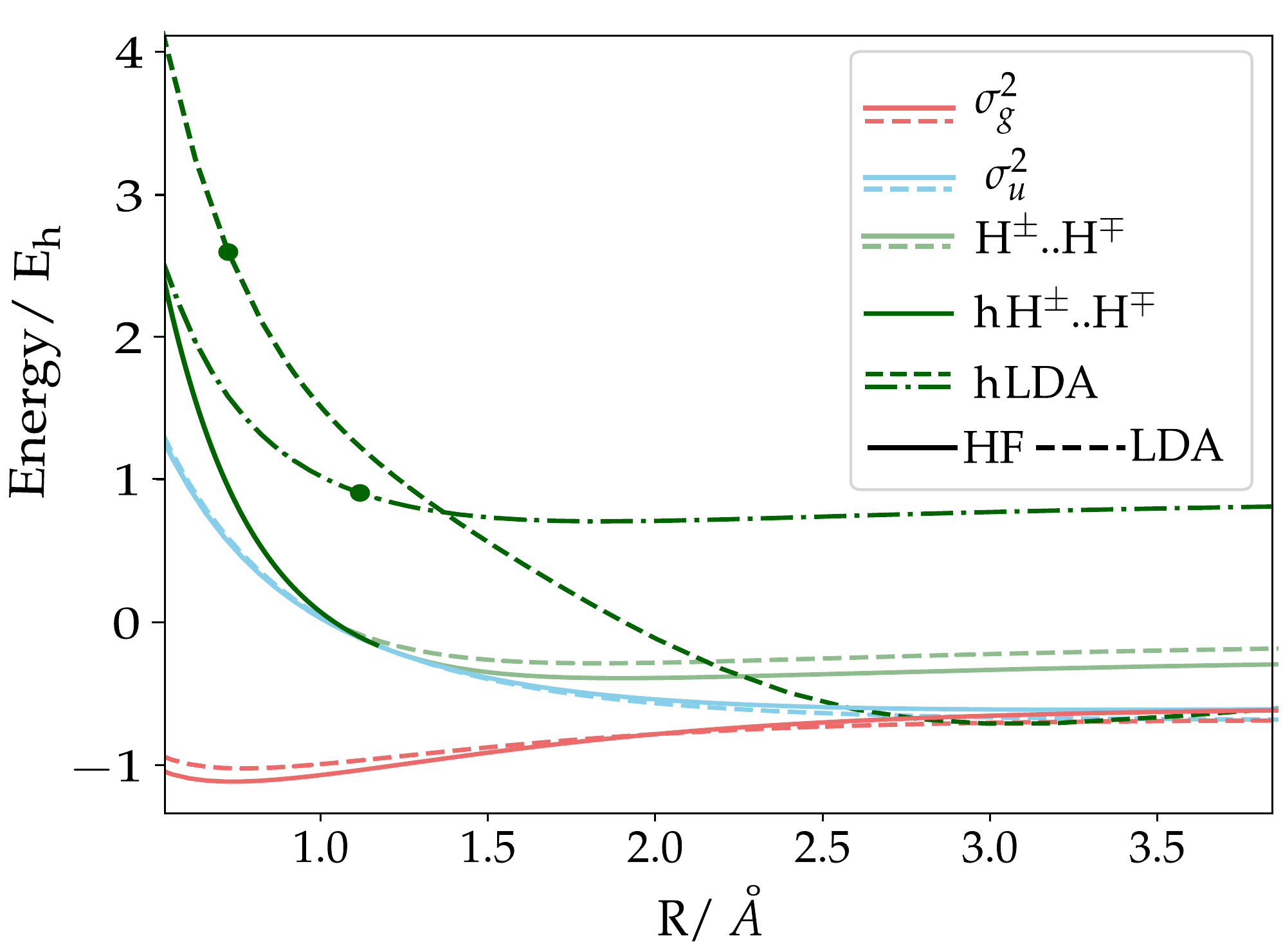}
\caption{Real contribution to the (holomorphic) electronic SCF energies of \ce{H2} along the bond dissociation coordinate using restricted HF theory (solid line) and restricted LDA-DFT (dashed line). The dark green dashed lines correspond to holomorphic LDA states which are obtained by tracing from holomorphic HF theory at the indicated points ($R=\SI{0.70}{\angstrom}$ and $R=\SI{1.10}{\angstrom}$).
}
\label{fig:hf_lda_overlay}
\end{figure}

The most famous example of SCF states coalescing occurs in the bond dissociation of $\ce{H2}$, as shown for the conventional restricted HF and LDA-DFT methods using the minimal STO-3G basis in Figure~\ref{fig:hf_lda_overlay}.
A total of $n=4$ stationary SCF states can be identified at large bond lengths,
labelled by their molecular structure as $\sigg^2$, $\sigu^2$ and the two degenerate ionic configurations H$^{\pm}$..H$^{\mp}$.\cite{Burton2018}  
In both LDA-DFT and HF, the $\sigu^2$ and H$^{\pm}$..H$^{\mp}$ states coalesce as the bond length is shortened.
The relative behaviour of the SCF states is the same for both LDA-DFT and HF, with the energy ordering and degeneracies of each state unchanged.
However, the absolute LDA-DFT energies, and therefore the location of the coalescence point, are shifted compared to the HF energies.
In HF theory, this coalescence point occurs at approximately $\SI{1.15}{\angstrom}$, whereas in LDA-DFT it is located at approximately $\SI{0.87}{\angstrom}$.

\begin{figure*}[htbp!]
\begin{subfigure}[b]{0.32\linewidth}
\begin{tikzpicture}
\draw (0,0) node[align=center] (plot) {%
\includegraphics[width=1\columnwidth]{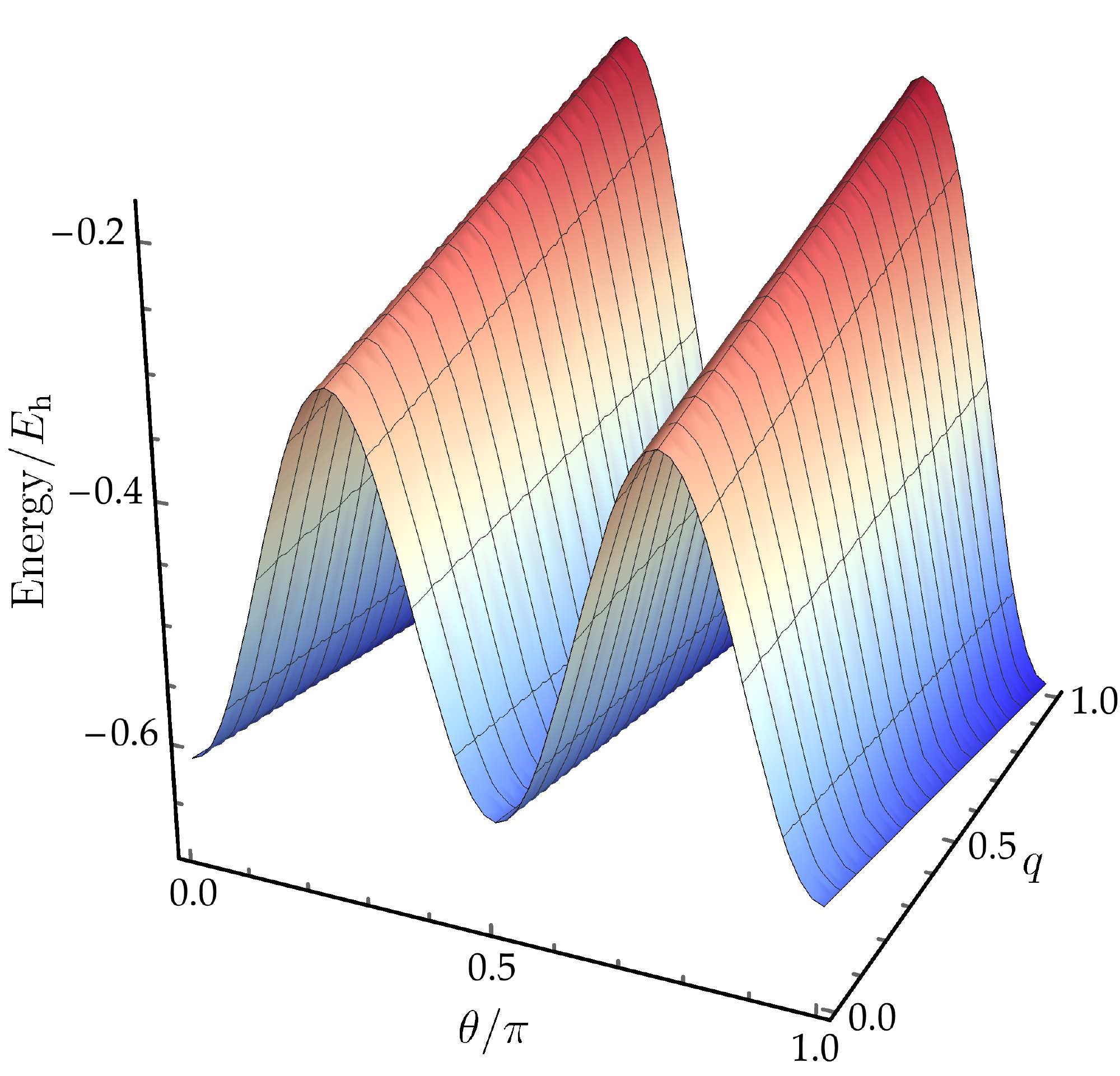}};
\draw (-1.5,-1.3) node[align=center] {$\sigg^2$};
\draw (0.5,2.8) node[align=center] { \ce{H^{+}-H^{-}} };
\draw (2.2,2.6) node[align=center] { \ce{H^{-}-H^{+}} };
\draw (0.0,-1.7) node[align=center] {$\sigu^2$};
\end{tikzpicture}
\subcaption{$R=\SI{4.00}{\angstrom}$}
\label{fig:uncoalesced}
\end{subfigure}
\begin{subfigure}[b]{0.32\linewidth}
\begin{tikzpicture}
\draw (0,0) node[align=center] (plot) {%
\includegraphics[width=1\columnwidth]{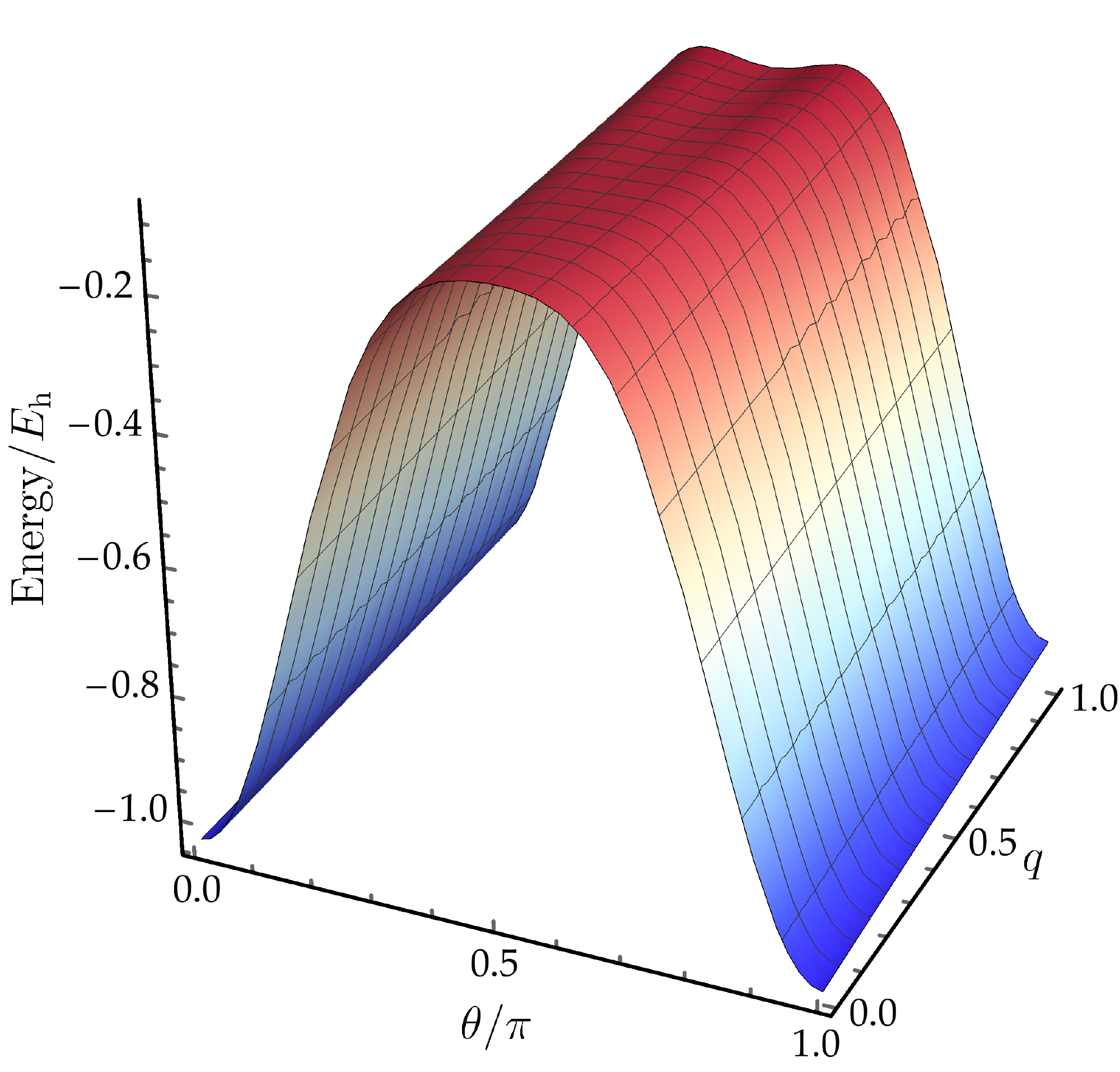}};
\draw (-1.2,-1.4) node[align=center] {$\sigg^2$};
\draw (0.5,2.8) node[align=center] { \ce{H^{+}-H^{-}} };
\draw (2.2,2.6) node[align=center] { \ce{H^{-}-H^{+}} };
\draw (-1.1,1.4) node[align=center] {$\sigu^2$};
\end{tikzpicture}
\subcaption{$R=\SI{1.10}{\angstrom}$}
\label{fig:inbetween}
\end{subfigure}
\begin{subfigure}[b]{0.32\linewidth}
\begin{tikzpicture}
\draw (0,0) node[align=center] (plot) {%
\includegraphics[width=1\columnwidth]{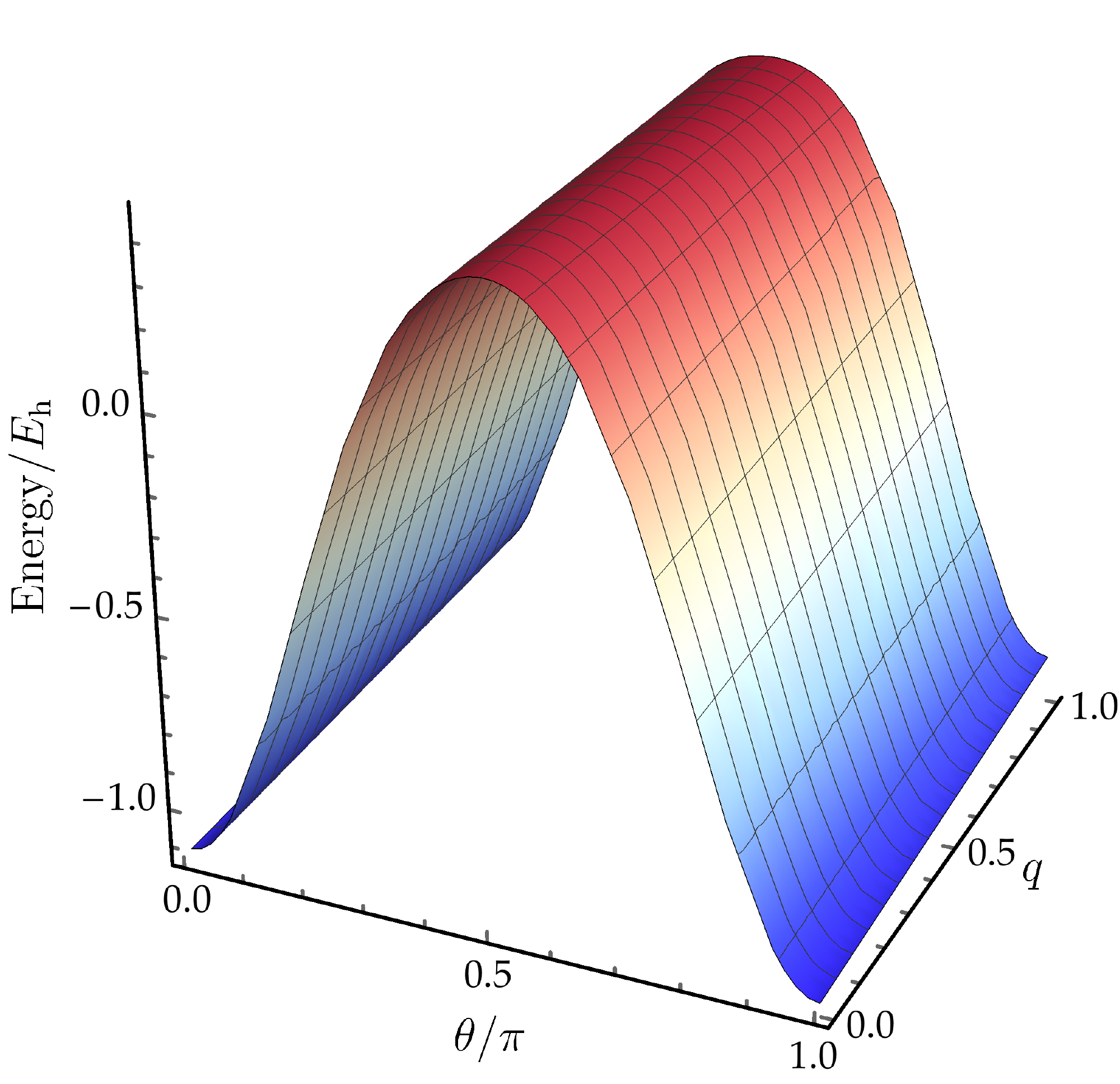}};
\draw (-1.2,-1.4) node[align=center] {$\sigg^2$};
\draw (1.2,2.8) node[align=center] {$\sigu^2$};
\end{tikzpicture}
\subcaption{$R=\SI{0.75}{\angstrom}$}
\label{fig:coalesced}
\end{subfigure}
\caption{%
SCF energy as a function of the rotation angle $\theta$ between the symmetry orbitals $\sigg$ and $\sigu$  and the scaling parameter $q$ between HF theory and LDA-DFT as described in Eq.~\eqref{convScaling}.
(\subref{fig:uncoalesced}) At a bond length of $R=\SI{4.00}{\angstrom}$, all four SCF solutions can be identified.
(\subref{fig:inbetween}) At a bond length of $R=\SI{1.10}{\angstrom}$, between the coalescence points of HF and LDA-DFT, the ionic states appear in the LDA-DFT limit ($q=1$) but have already coalesced in a HF framework ($q=0$).
(\subref{fig:coalesced}) In the equilibrium regime at a bond length of $R=\SI{0.75}{\angstrom}$, only two stationary points are observed and the ionic SCF states have vanished in both LDA-DFT and HF.}
\label{rotations}
\end{figure*}

The closed-shell SCF wave function for \ce{H2} contains only one doubly occupied spatial orbital $\phi\qty(\br)$ which can be expanded in terms of the (real orthogonal) MO basis using the rotation angle $\theta$ as
\begin{equation}
\phi\qty(\br) = \sigg\qty(\br) \cos\theta + \sigu\qty(\br) \sin\theta
\label{eq:OccOrb}
\end{equation}
The coalescence of states of \ce{H2} for a linear interpolation between HF and LDA-DFT can be visualised by parametrising the SCF energy surface in terms of the exchange-correlation scaling $q$ and the orbital rotation angle $\theta$, as shown in Figure~\ref{rotations}.
For a given scaling $q$ or bond length $R$, the number of real SCF states corresponds to the number of stationary points with respect to $\theta$.
As expected,\cite{Burton2018} there are four stationary points for both HF and LDA-DFT at large bond lengths (Figure~\ref{fig:uncoalesced}), and only two stationary points at the equilibrium geometry while the ionic solutions having vanished at the coalescence point (Figure~\ref{fig:coalesced}) .

The bonding $\sigg^2$ and anti-bonding $\sigu^2$ states can be identified for all bond lengths in both HF and LDA-DFT. 
In contrast, between the HF and LDA-DFT Coulson--Fischer points, the ionic states that have disappeared in the HF framework reappear in the LDA-DFT one (Figure~\ref{fig:inbetween}).
This observation is surprising since it is the \emph{opposite} scenario to the model electron transfer, where the symmetry-broken SCF states existed in the HF case but not LDA-DFT.
We believe that 
different types of SCF symmetry breaking (eg.\ singlet or triplet instabilities) may have distinct coalescence patterns upon scaling between HF and DFT, in turn suggesting fundamental differences in the types of symmetry breaking using various functionals.
At bond lengths shorter than the coalescence point, it it is known that h-HF solutions continue to exist with complex-valued orbital coefficients.\cite{Burton2018}
We expect our analytic h-DFT approximation to allow both real- and complex-valued h-HF stationary states to be mapped  onto the h-LDA states.
 
Like h-HF theory, we require the density-fitting h-DFT energy to retain the conventional (real) SCF solutions when they exist. 
To verify our h-DFT approach, we therefore applied the density-fitting method to study paths between the SCF states that are real in both HF and LDA-DFT.
We rewrite the nuclear-attraction energy $E_\text{nuc}$ in terms of the density-fitting basis to ensure that the HF energy is also dependent on the fitting parameters $f$,
allowing us to continuously link holomorphic HF theory and DFT.
Comparing the electronic energies for the $\sigg^2$ and $\sigu^2$ states from the holomorphic density fitting and a reference calculation using \textsc{Q-Chem~5.2}\cite{Q-Chem} demonstrates a sub-$\text{mE}_{\text{h}}$ agreement for both HF and LDA-DFT calculations, as shown at $R=\SI{6.35}{\angstrom}$ in Table~\ref{12_bohr}.
However, although \ce{H2} has exactly four true h-RHF solutions using a minimal basis,\cite{Burton2018}
solving for all RHF states in the density-fitting implementation leads to more than four solutions.
Many of these solutions have the same energies as the exact h-HF solutions, and we believe that the additional solutions arise from implicitly allowing different cubed-roots of the electron density.
Clearly only the real cube-root of the electron density provides a physical solution since the additional SCF solutions using the density-fitting equations do not link to the real-valued HF energies where they exist.
We therefore believe that these additional solutions are mathematical artefacts, and safely be ignored.

\begin{table}[h!]
\begin{ruledtabular}
\begin{tabular}{ld{2.6}d{2.6}}
                      & \head{$E\qty(\sigg^2)$ } & \head{$E \qty(\sigu^2)$}
\Tstrut\Bstrut\\
\hline\Tstrut
\textbf{HF ($q=0.0$)}             &              &           
\\
Holomorphic Density Fitting&  -0.587464   & -0.241891 
\\
Reference (\textsc{Q-Chem})       &  -0.587531   & -0.241891 
\\
\hline\Tstrut
\textbf{LDA-DFT ($q=1.0$)}        &              &           
\\
Holomorphic Density Fitting& -0.685653    & -0.131509 
\\
Reference (\textsc{Q-Chem})       & -0.685748    & -0.131498 
\\
\end{tabular}
\end{ruledtabular}
\caption{%
HF and LDA-DFT energies of the conventional SCF states computed using the holomorphic density fitting approach and a reference calculation using \textsc{Q-Chem~5.2}\cite{Q-Chem} are found to agree with sub-$\text{mE}_\text{h}$ accuracy at $R = \SI{6.35}{\angstrom}$.
All energies are given in atomic units of Hartrees ($\Eh$).
\label{12_bohr}
}
\end{table}



\begin{figure*}[htbp]
	\centering
	\begin{subfigure}[b]{.8\columnwidth}
		\centering
		\hspace{-3em}\includegraphics[width=\columnwidth]{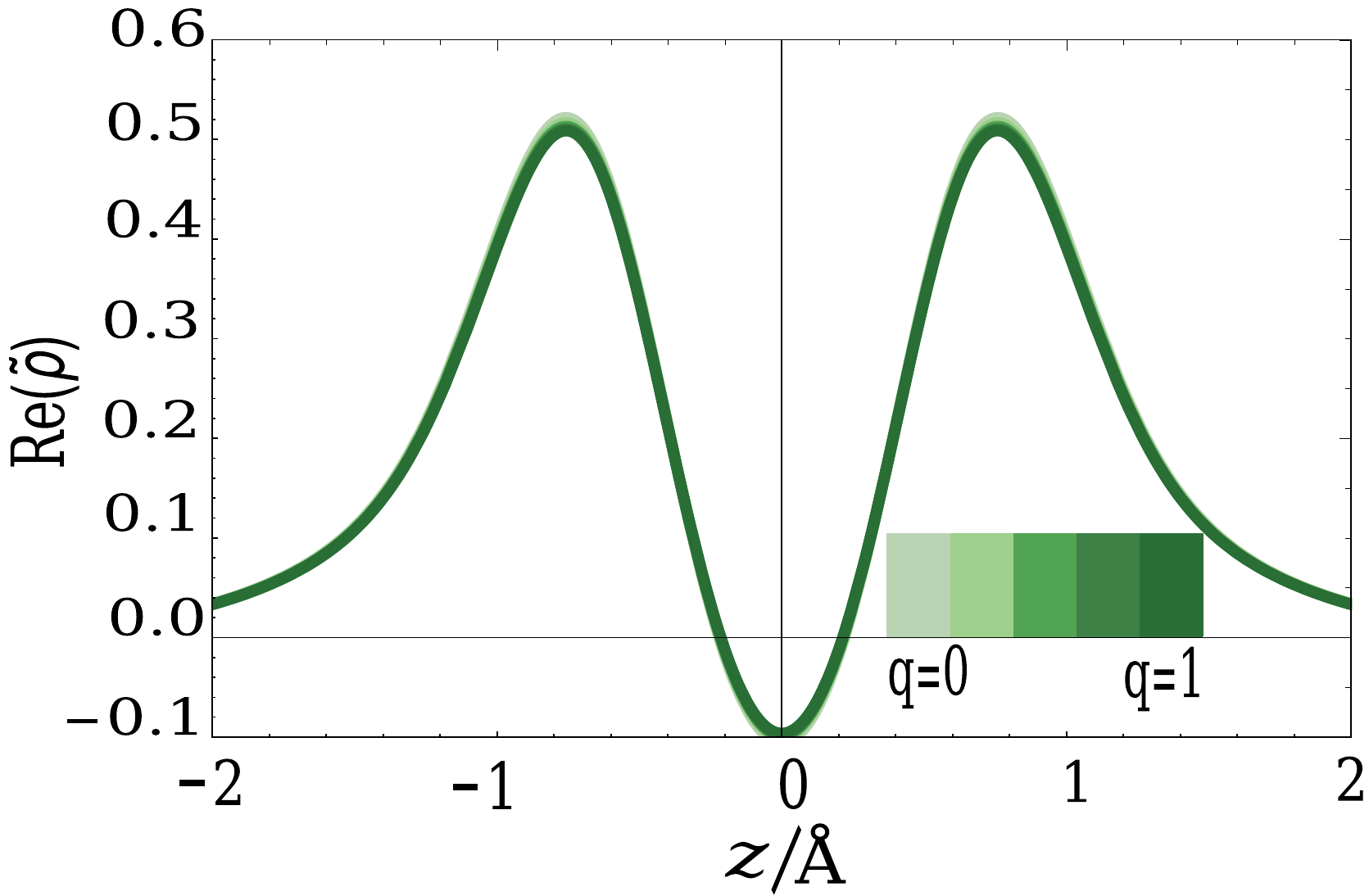}
		\subcaption{}    
		\label{fig:real_holoDens_R070}
	\end{subfigure}
	\begin{subfigure}[b]{.8\columnwidth}  
		\centering 
		\hspace{-3em}\includegraphics[width=\columnwidth]{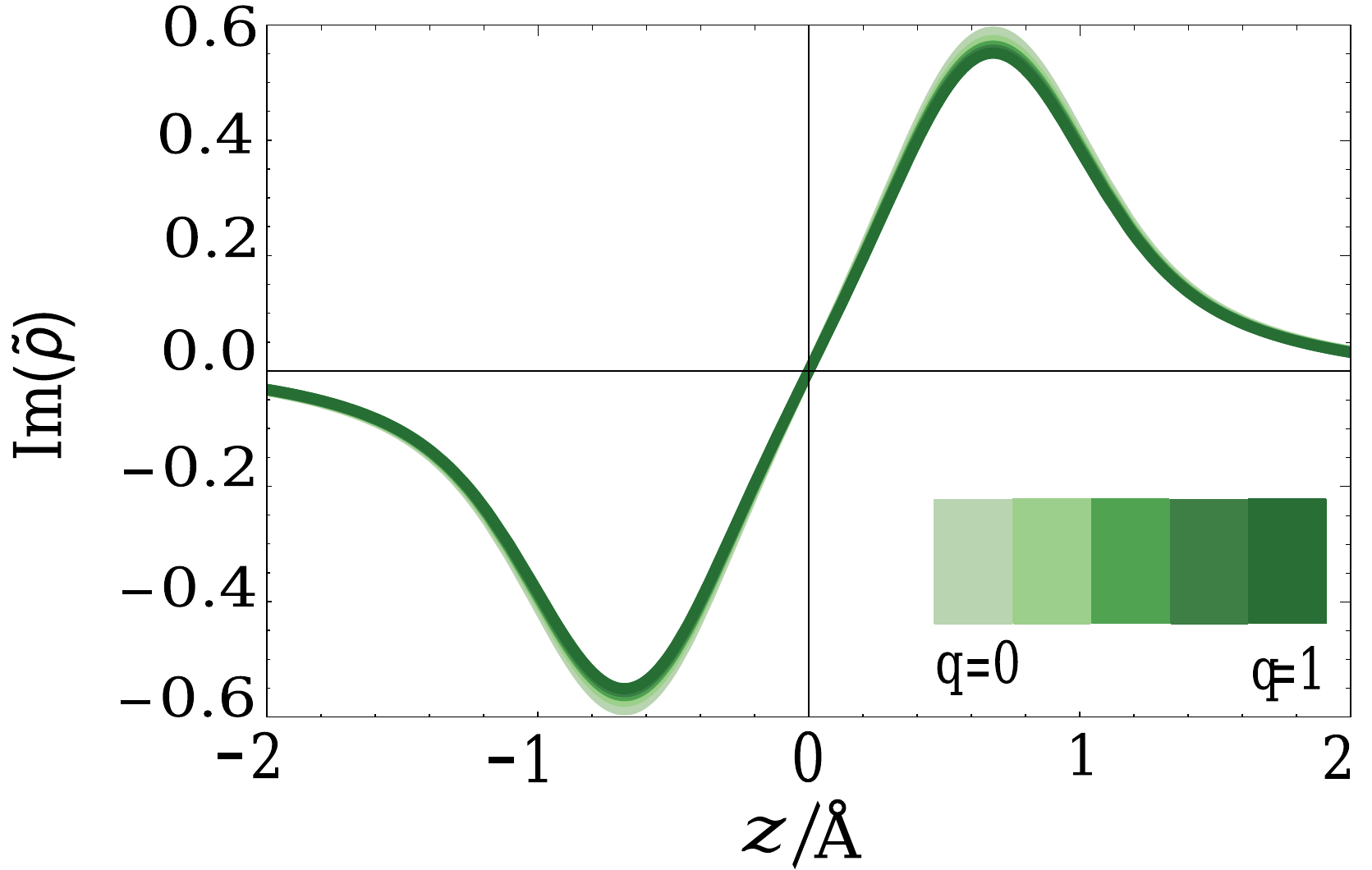}
		\subcaption{}
		\label{fig:imag_holoDens_R070}
	\end{subfigure}
	\begin{subfigure}[b]{.8\columnwidth}
		\centering
		\hspace{-3em}\includegraphics[width=\textwidth]{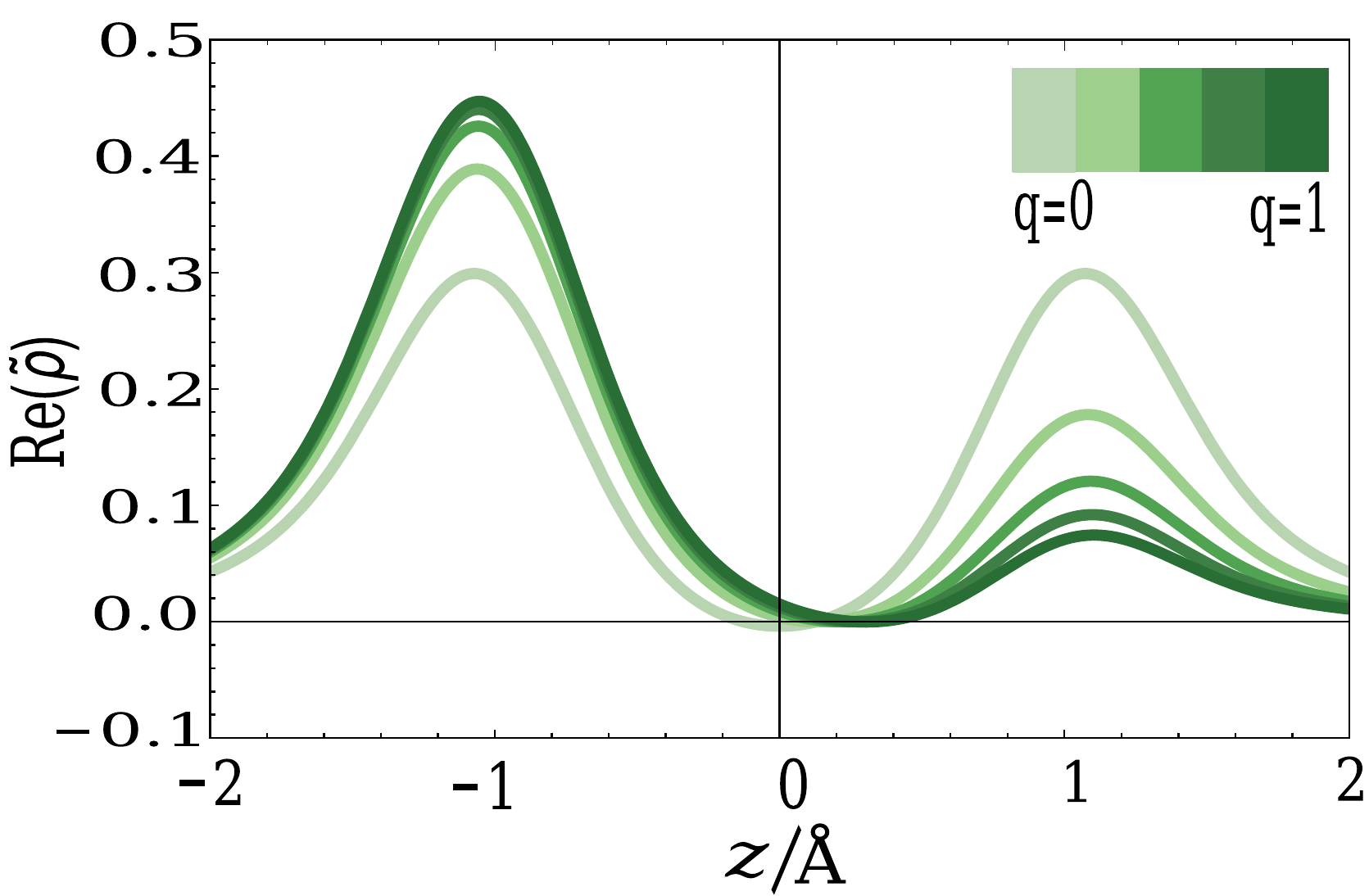}
		\subcaption{}    
		\label{fig:holo_cdens_r_b}
	\end{subfigure}
	\begin{subfigure}[b]{.8\columnwidth}  
		\centering 
		\hspace{-3em}\includegraphics[width=\columnwidth]{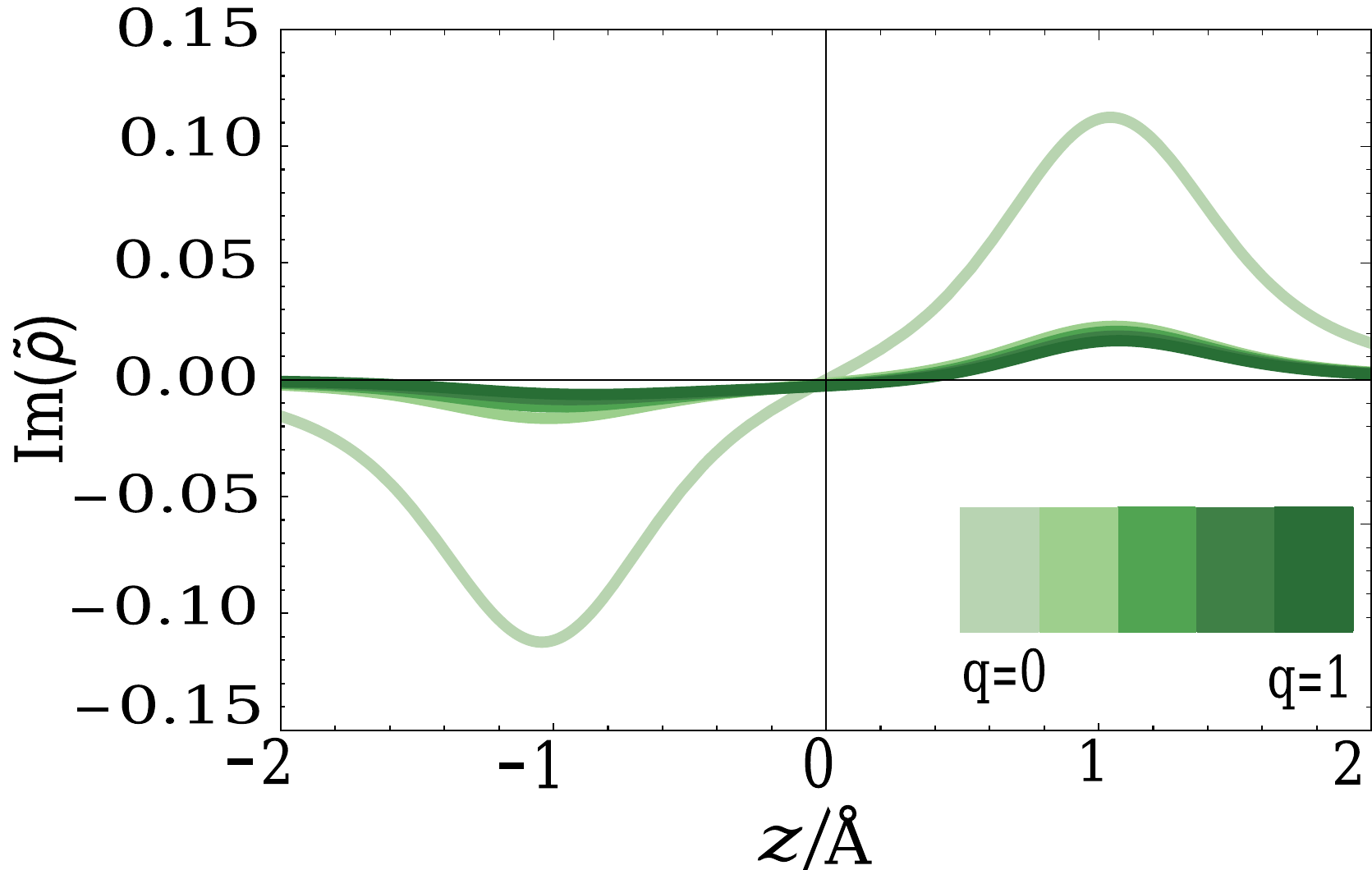}
		\subcaption{}
		\label{fig:holo_cdens_i_b}
	\end{subfigure}
    \begin{subfigure}[b]{.8\columnwidth}
		\centering
		\hspace{-3em}\includegraphics[width=\columnwidth]{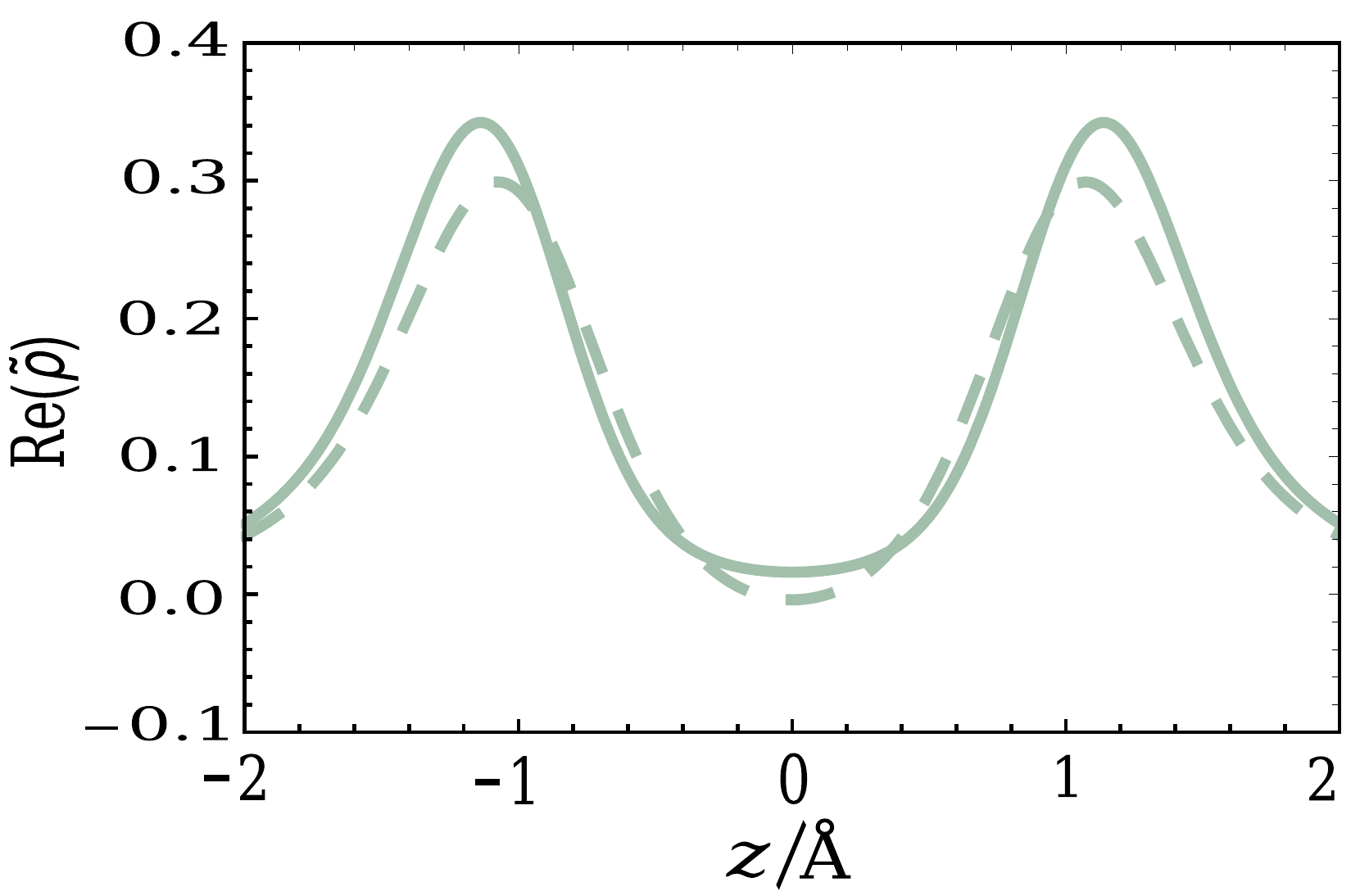}
		\subcaption{}    
		\label{fig:fitting_r}
	\end{subfigure}
	\begin{subfigure}[b]{.8\columnwidth}  
		\centering 
		\hspace{-3em}\includegraphics[width=\columnwidth]{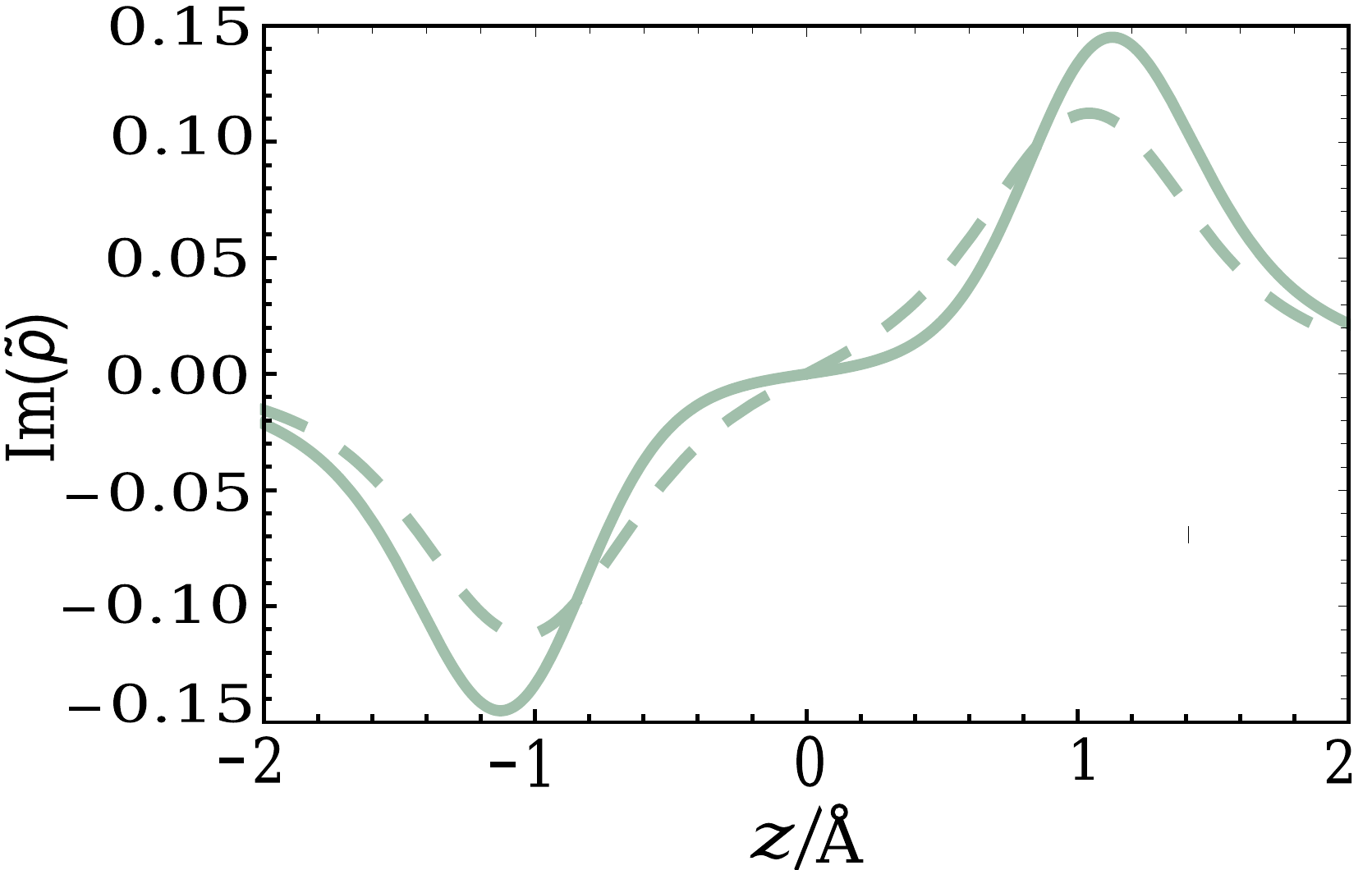}
		\subcaption{}
		\label{fig:fitting_i}
	\end{subfigure}
	\caption{Real (\subref{fig:real_holoDens_R070}) and imaginary (\subref{fig:imag_holoDens_R070}) components of the holomorphic electron density $\tilde{\rho}\qty(\br)$ of the h-SCF state for different scaling factors $q$ at a bond length of $R=\SI{0.70}{\angstrom}$. 
    Real (\subref{fig:holo_cdens_r_b}) and imaginary (\subref{fig:holo_cdens_i_b}) components of the holomorphic electron density $\tilde{\rho}\qty(\br)$ of the ionic state h-SCF  for different scaling factors $q$ at a bond length of $R=\SI{1.10}{\angstrom}$. 
    Real (\subref{fig:fitting_r}) and imaginary (\subref{fig:fitting_i}) components of the holomorphic electron density $\tilde{\rho}\qty(\br)$ of the h-RHF state at a bond length of $R=\SI{1.10}{\angstrom}$ calculated from the AO basis $\ket{\Psi}$ (solid line) and the density fitting basis $\ket{\xi}$ (dashed line).
	}
	\label{adiabat_change}
\end{figure*}

Beyond the coalescence point in HF theory, we can always successfully identify the complex h-HF extension of the ionic state using the density fitting method. 
In principle, the holomorphic density-fitting method therefore allows both real and holomorphic SCF states to be continuously mapped from HF theory to LDA-DFT.
In practice, however, the success of the density fitting depends strongly on the size and choice of the density fitting basis.

To identify LDA extensions of the h-HF state, we first traced the h-HF solution using the density fitting across at $\SI{0.70}{\angstrom}$, where we expect the complex h-HF state to evolve continuously into the complex h-LDA solution. 
While there are three degenerate density-fitting solutions in the h-HF framework that provide a real holomorphic energy which matches with the conventional h-HF implementation,\cite{Burton2018} only one of these solutions retains a meaningful holomorphic energy when traced along to h-LDA-DFT, corresponding to the correct cube-root density.
Tracking the holomorphic ionic state from HF to LDA-DFT by relaxing the wave function at each scaling value, we find that the MO coefficients, and in turn the holomorphic electron density $\tilde{\rho}\qty(\br)$, do not change significantly between the two functionals,  as shown in Figures~\ref{fig:real_holoDens_R070} and \ref{fig:imag_holoDens_R070}.
Within the h-LDA-DFT framework, we then traced this solution along bond dissociation coordinate, giving the binding curve shown in Figure~\ref{fig:hf_lda_overlay}.
Surprisingly, the energy of this state did not converge onto the energy of the conventional ionic LDA-state at the coalescence point ($R=\SI{0.87}{\angstrom}$), as would be expected for a complex-analytic extension of the real ionic state.

To understand this effect, we then tracked the h-HF solution from HF to LDA at a bond length between the coalescence points of the two potentials ($R=\SI{1.10}{\angstrom}$), where the ionic state has already coalesced in HF theory but remains a real stationary point in the LDA-DFT framework.
For this bond length, there is a large rearrangement in the holomorphic electron density and, contrary to what we expect, we recover a complex-valued holomorphic SCF solution in h-LDA-DFT as well.
The real part of the holomorphic density (Figure \ref{fig:holo_cdens_r_b}) retains symmetry-broken ionic character in the limit of a LDA-DFT calculation, as expected since the ionic LDA-DFT state has not yet coalesced, but reduces to a more symmetric distribution in the HF case.
The imaginary component of the holomorphic electron density $\tilde{\rho}\qty(\br)$ however remains non-zero when tracking the holomorphic SCF state from HF to LDA-DFT, even though the ionic states exist in LDA-DFT (Figure~\ref{fig:holo_cdens_i_b}). 
This state also yields a complex-valued holomorphic energy rather than recovering the real-valued ionic LDA-DFT energy that would be expected.
Moreover, its energy and electron density do not match the h-LDA state traced from shorter bond lengths, and following this new solution along the binding curve
reveals that it also does not converge onto the real LDA ionic state (see Figure \ref{fig:hf_lda_overlay}).
%
%
%

Clearly, the density fitting method is capable of yielding complex h-LDA states, but we have not been able to find a unique solution that corresponds to a complex-analytic extension of the real ionic state.
The current density-fitting implementation's failure to recover the holomorphic ionic state can be attributed to the fitting quality for the h-HF electron density.
Figures \ref{fig:fitting_r} and \ref{fig:fitting_i} compare the real and imaginary part of the holomorphic electron density calculated using the MO coefficients and the density-fitting coefficients.
Both real and imaginary parts are not fitted particularly well using the Gaussian-like fitting basis, indicating a need for a larger fitting basis, potentially with a different functional form.
While the density-fitting approach is conceptually exact, and recovers the desired polynomial electronic energy functional, it is too numerically challenging to apply in practice.

\subsection{Holomorphic Kohn--Sham Theory}
\label{sec:KohnSham}

\begin{figure*}[tbhp]
\begin{subfigure}[b]{0.49\linewidth}
\includegraphics[width=\textwidth, trim=0pt 52pt 0pt 0pt, clip]{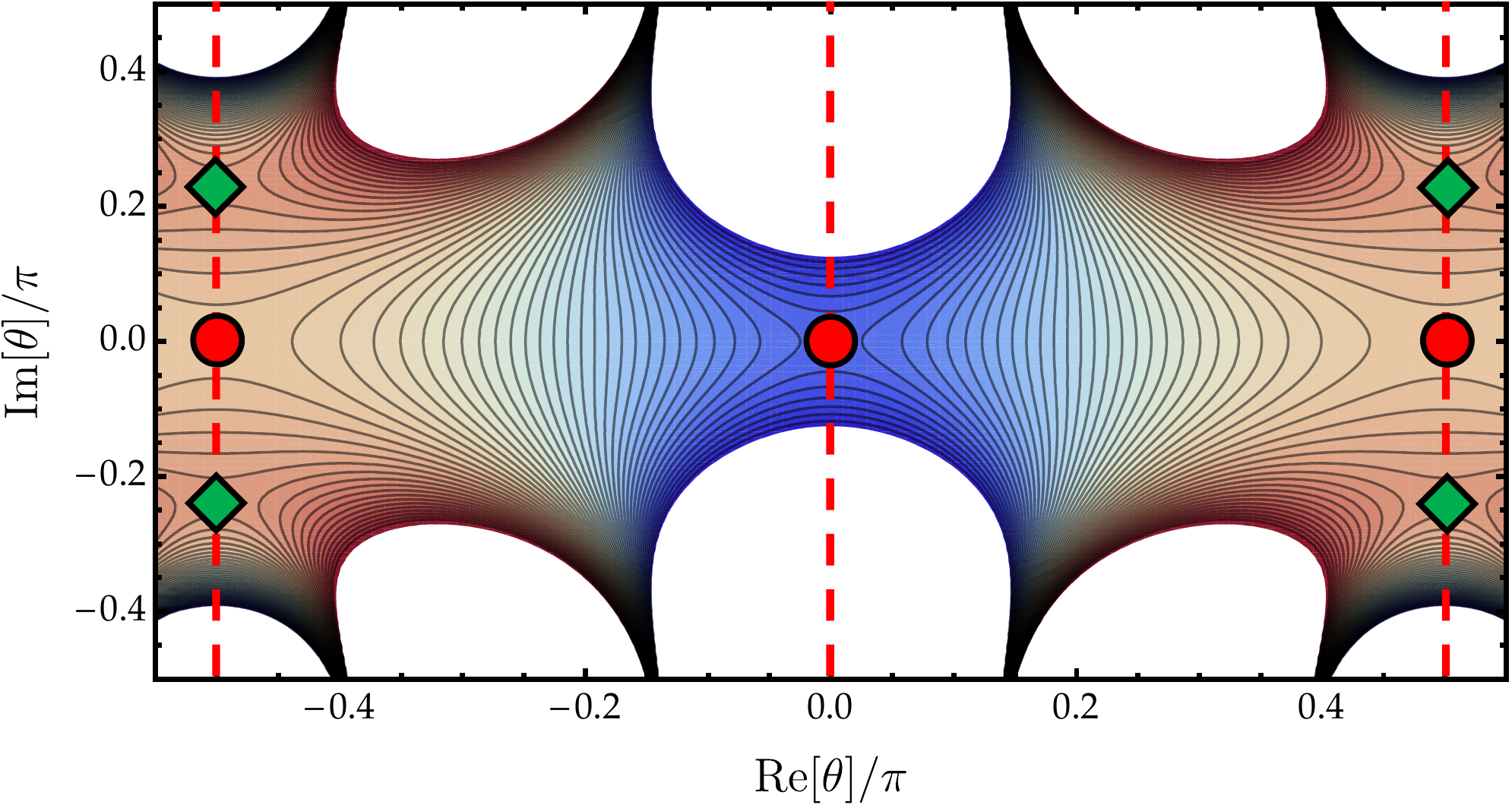}\\
\includegraphics[width=\textwidth, trim=0pt 0pt 0pt 1pt, clip]{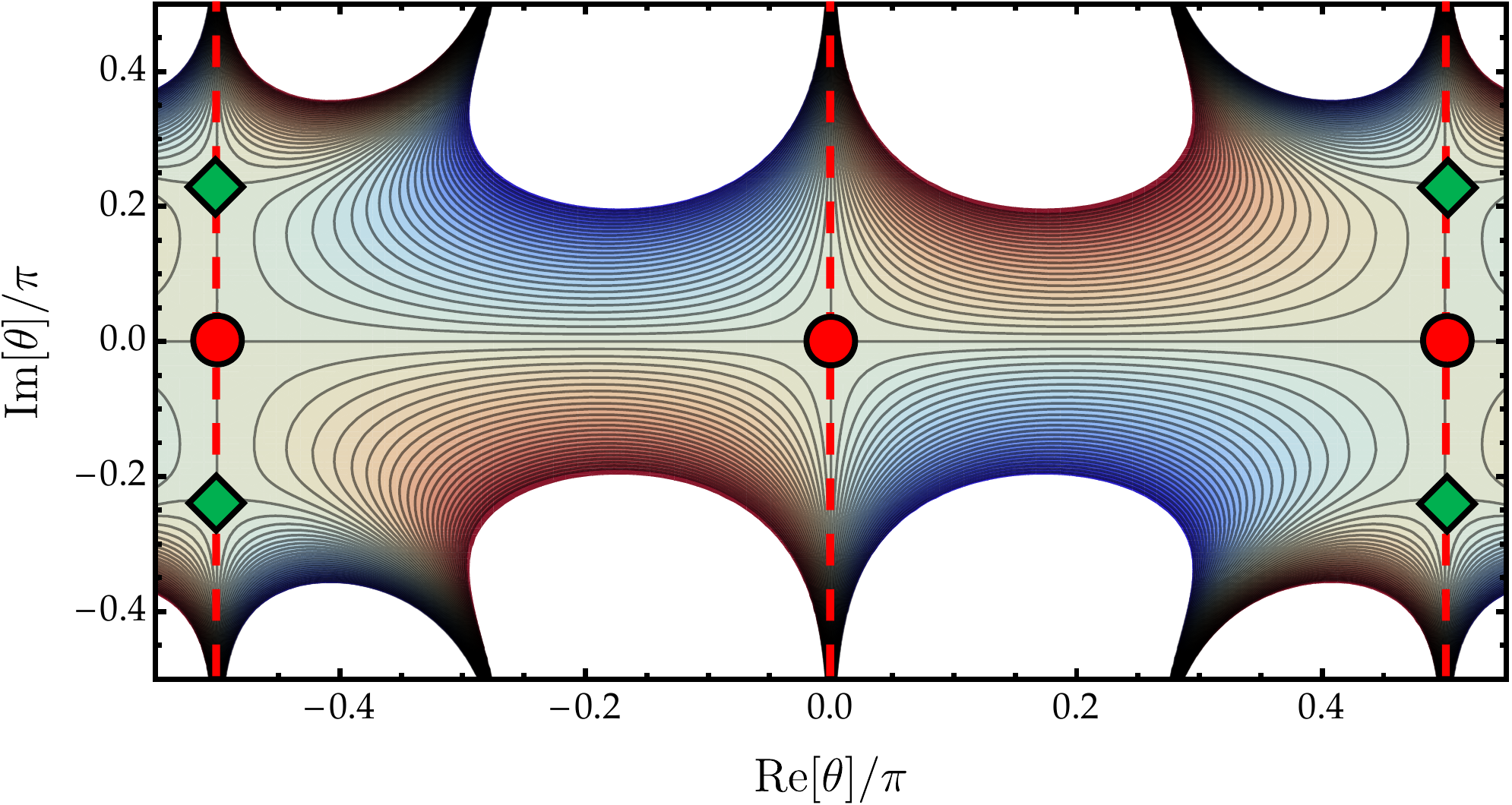} 
\subcaption{Hartree--Fock}    
\label{subfig:SurfHF}
\end{subfigure}
\hfill
\begin{subfigure}[b]{0.49\linewidth}
\includegraphics[width=\textwidth, trim=0pt 52pt 0pt 0pt, clip]{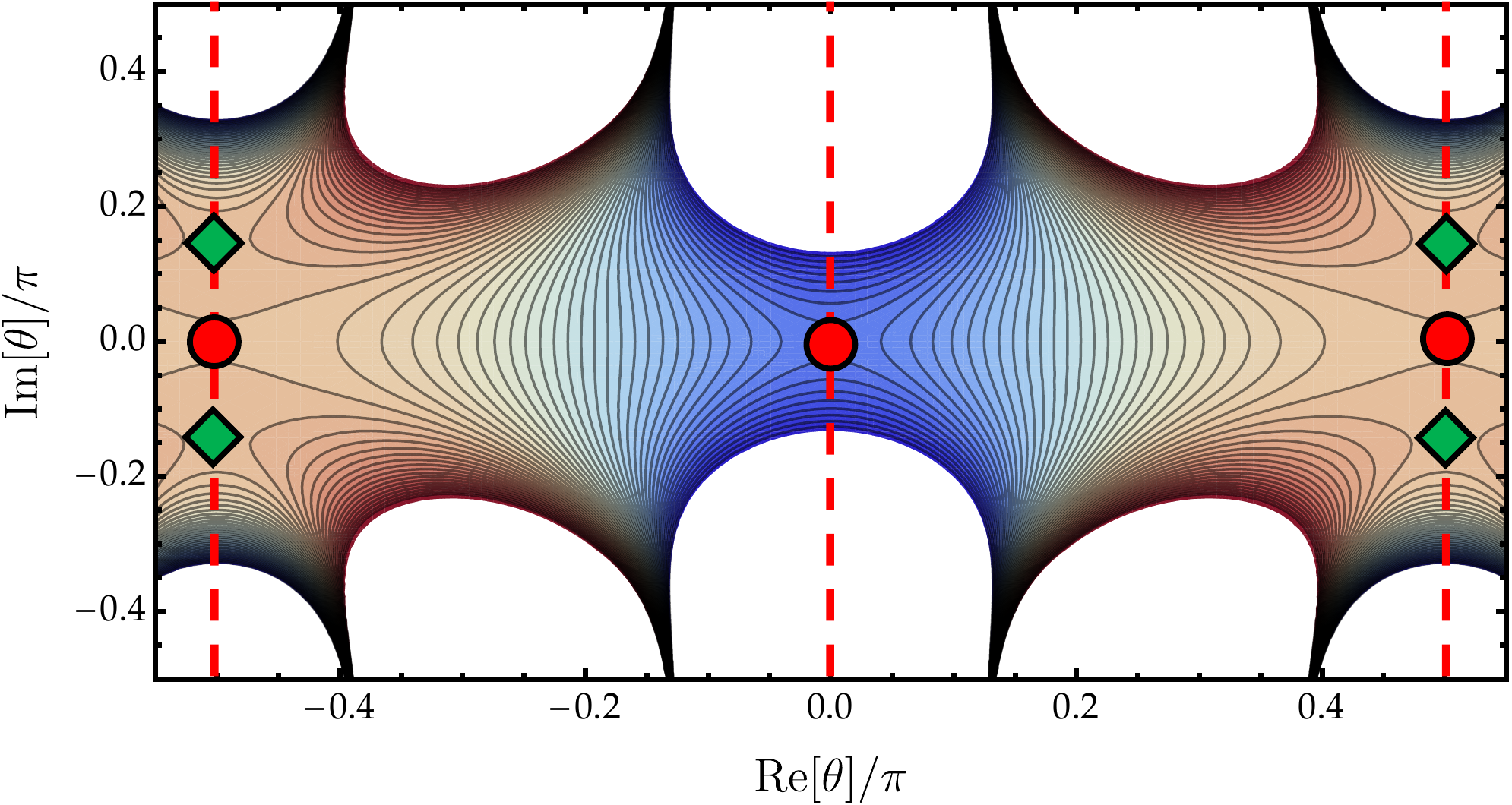} \\
\includegraphics[width=\textwidth, trim=0pt 0pt 0pt 1pt, clip]{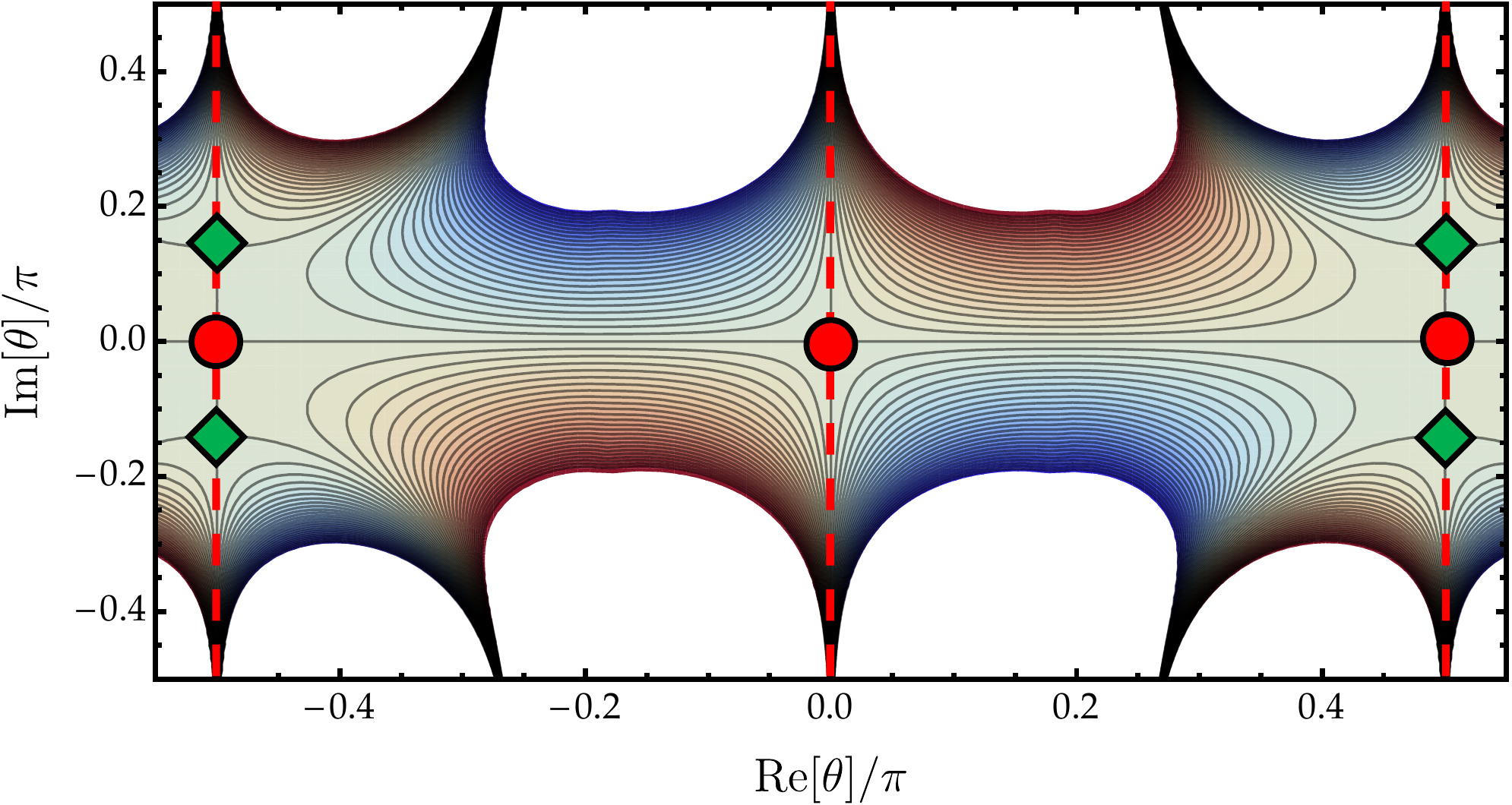} 
\subcaption{LDA}    
\label{subfig:SurfLDA}
\end{subfigure}
\caption{Comparison of the real (top) and imaginary (bottom) components of the (\subref{subfig:SurfHF}) h-HF and  (\subref{subfig:SurfLDA}) h-LDA energy surfaces for a restricted SCF wave function parameterised by the single occupied orbital~\eqref{eq:OccOrb}. 
The parity-time symmetric line (dashed red) has purely real h-SCF energies. 
Complex-valued h-HF and h-LDA solutions are indicated by a green diamond, and occur along the parity-time symmetric line for both potentials.
The symmetry-pure $\sigg^2$ and $\sigu^2$ solutions correspond to the red circles at $\theta=0$~and~$\pm \pi / 2$ respectively.
}
\label{fig:SCFsurfaces}
\end{figure*}

To overcome the numerical issues of the density fitting approach, we also considered a na\"{i}ve implementation of h-DFT by removing any complex conjugation 
of the MO coefficients from the SCF energy function.
Following the philosophy of Ref.~\onlinecite{Burton2016}, we simply defined the h-LDA functional using the holomorphic density defined in Eq.~\eqref{eq:HoloDensity}.
The corresponding restricted SCF energy surfaces can then be visualised as a function of the complex rotation angle $\theta$ that defines
the single occupied orbital Eq.~\eqref{eq:OccOrb}, as shown in Figure~\ref{fig:SCFsurfaces}.
We find a remarkable similarity between the topology of the h-HF and h-LDA energy surfaces for all complex values of $\theta$.
The number of stationary points is the same for both potentials, with two solutions along the real axis corresponding to the $\sigg^2$ and $\sigu^2$ states, 
and two stationary points with complex-valued orbitals.
These complex-valued stationary points correspond to the holomorphic extensions of the ionic SCF states, 
confirming that complex holomorphic extensions can also be identified for solutions that disappear  using the LDA functional. 

Analogously to the HF case, the complex h-LDA solutions occur along a line of strictly real energies defined by $\theta = \frac{\pi}{2} + \vartheta \mathrm{i}$, 
for $\vartheta \in \mathbb{R}$ (red dashed line in Figure~\ref{fig:SCFsurfaces}).
In HF theory, this line has recently been shown to correspond with the conservation of parity-time symmetry in the molecular wave function,\cite{Burton2019b} providing a weaker 
condition than Hermiticity for ensuring real electronic energies.\cite{Bender1998}
Our observations therefore provide the first evidence of parity-time symmetry in the DFT framework, paving the way for novel
applications of this new symmetry across single-particle approximations.

Encouraged by the existence of complex stationary points on the h-LDA energy surface, we implemented a holomorphic KS (h-KS) approach to iteratively identify h-DFT solutions. 
Following the original h-HF SCF approach,\cite{Burton2016} the closed-shell holomorphic Fock matrix for the LDA functional is defined as 
\begin{equation}
\tilde{F}_{\mu \nu} = h_{\mu \nu} + j_{\mu \nu} + \tilde{F}^{\text{LDA}}_{\mu \nu},
\label{eq:HoloFock}
\end{equation}
where the exchange-correlation contribution\cite{Pople1992} to the Fock matrix $\tilde{F}^{\text{LDA}}_{\mu \nu}$ is defined in terms of the holomorphic density as
\begin{equation}
\tilde{F}^{\text{LDA}}_{\mu \nu} = -\qty(\frac{6}{\pi})^{\frac{1}{3}} \int \tilde{\rho}\qty(\br)^{\frac{1}{3}} \mathrm{d}^3 \br.
\end{equation}
The h-KS algorithm then proceeds in a analogous manner to the h-HF approach,\cite{Burton2016} with new occupied orbitals
on each iteration selected using a complex-symmetric extension of the maximum overlap method,\cite{Gilbert2008} 
and convergence accelerated using the DIIS (direct inversion in the iterative subspace) extrapolation scheme.\cite{Pulay}
This approach can be trivially extended to a spin-unrestricted formalism by defining $\alpha$ and $\beta$ Fock matrices in terms of the corresponding $\alpha$ and $\beta$ 
holomorphic densities.
The use of the complex-symmetric holomorphic density matrix is closely related to non-Hermitian extensions of KS-DFT developed to describe
metastable resonance states,\cite{Ernzerhof2006} although here we do not introduce any complex absorbing potential.

\begin{figure}[b!]
\includegraphics[width=\linewidth]{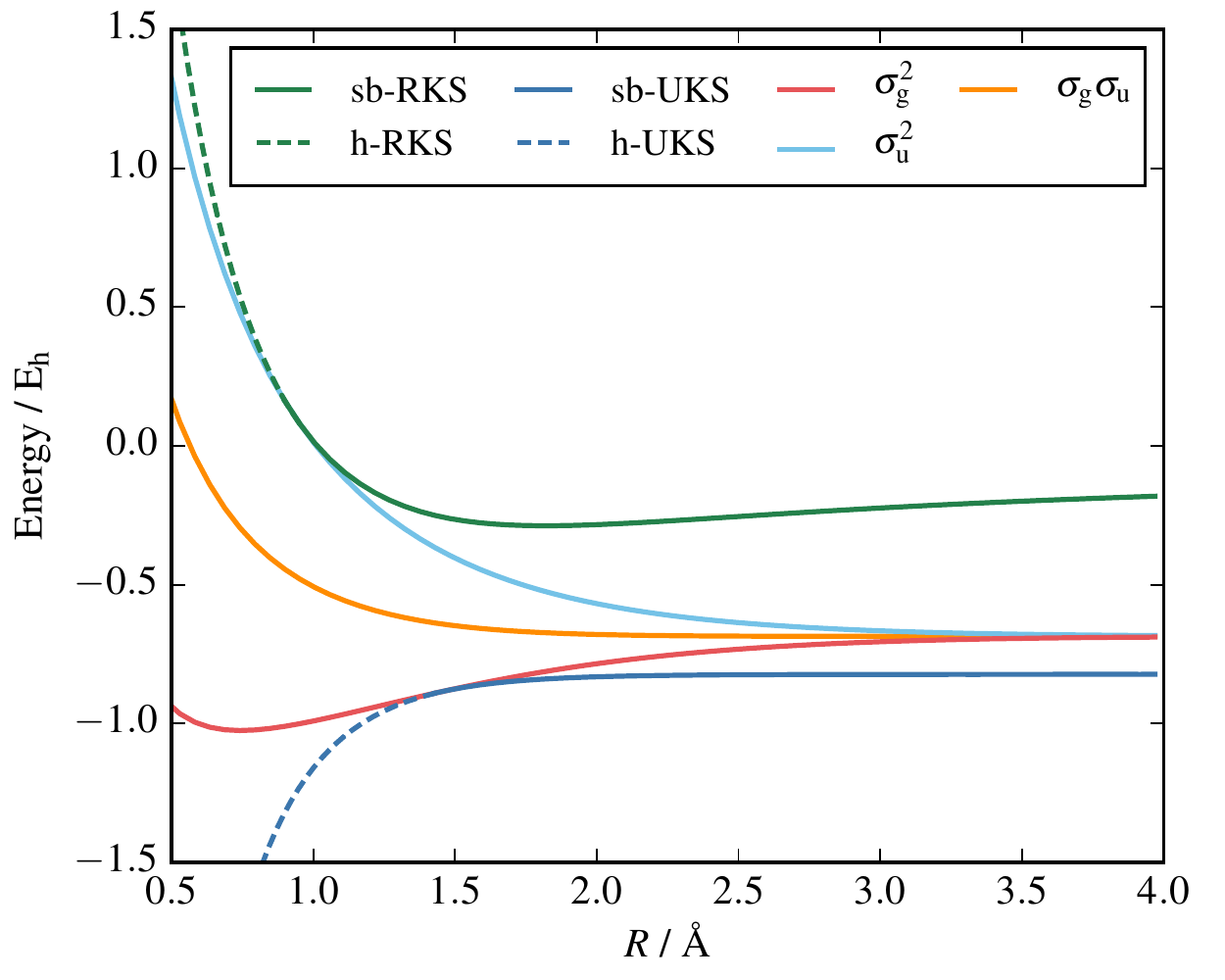}
\caption{Binding curves for the eight KS-LDA solutions. When the real symmetry-broken (sb-) RKS and UKS solutions disappear, complex-valued h-RKS and h-UKS solutions continue to exist (dashed lines).}
\label{fig:hKS_bindingCurve}
\end{figure}

Taking initial guesses for the optimal orbital coefficients from the h-LDA energy surfaces in Figure~\ref{fig:SCFsurfaces}, we have 
identified a total of four self-consistent h-RKS solutions  at $R=\SI{0.70}{\angstrom}$. 
Two of these solutions have real orbital coefficients, corresponding to the $\sigg^2$ and $\sigu^2$ solutions, while the remaining two have complex-valued orbital
coefficients and occur as the h-RKS complex conjugate pair. 
In addition, we identified a further four holomorphic unrestricted KS (h-UKS) solutions, including two with complex-valued orbital 
coefficients that correspond to the h-UHF states seen in previous studies.\cite{Burton2016}
All eight solutions can then be traced along the full binding curve by using the converged coefficients at one geometry as the 
initial guess at the next geometry, as shown in Figure~\ref{fig:hKS_bindingCurve}.
Numerical values for the orbital coefficients and h-LDA energies of each stationary point at bond lengths of $R = 0.70$, $1.10$, and $\SI{4.00}{\angstrom}$ are 
available in the Supporting Information.
Crucially, we find that the h-RKS and h-UKS solutions emerge from the coalescence points at which the real symmetry-broken ionic (sb-RKS) or diradical (sb-UKS) solutions disappear respectively.
These self-consistent h-KS solutions therefore provide the rigorous complex-analytic extensions of real KS solutions that disappear as the molecular structure changes.

\begin{figure*}[htbp]
	\centering
	\begin{subfigure}[b]{.8\columnwidth}
		\centering
		\hspace{-3em}\includegraphics[width=\columnwidth]{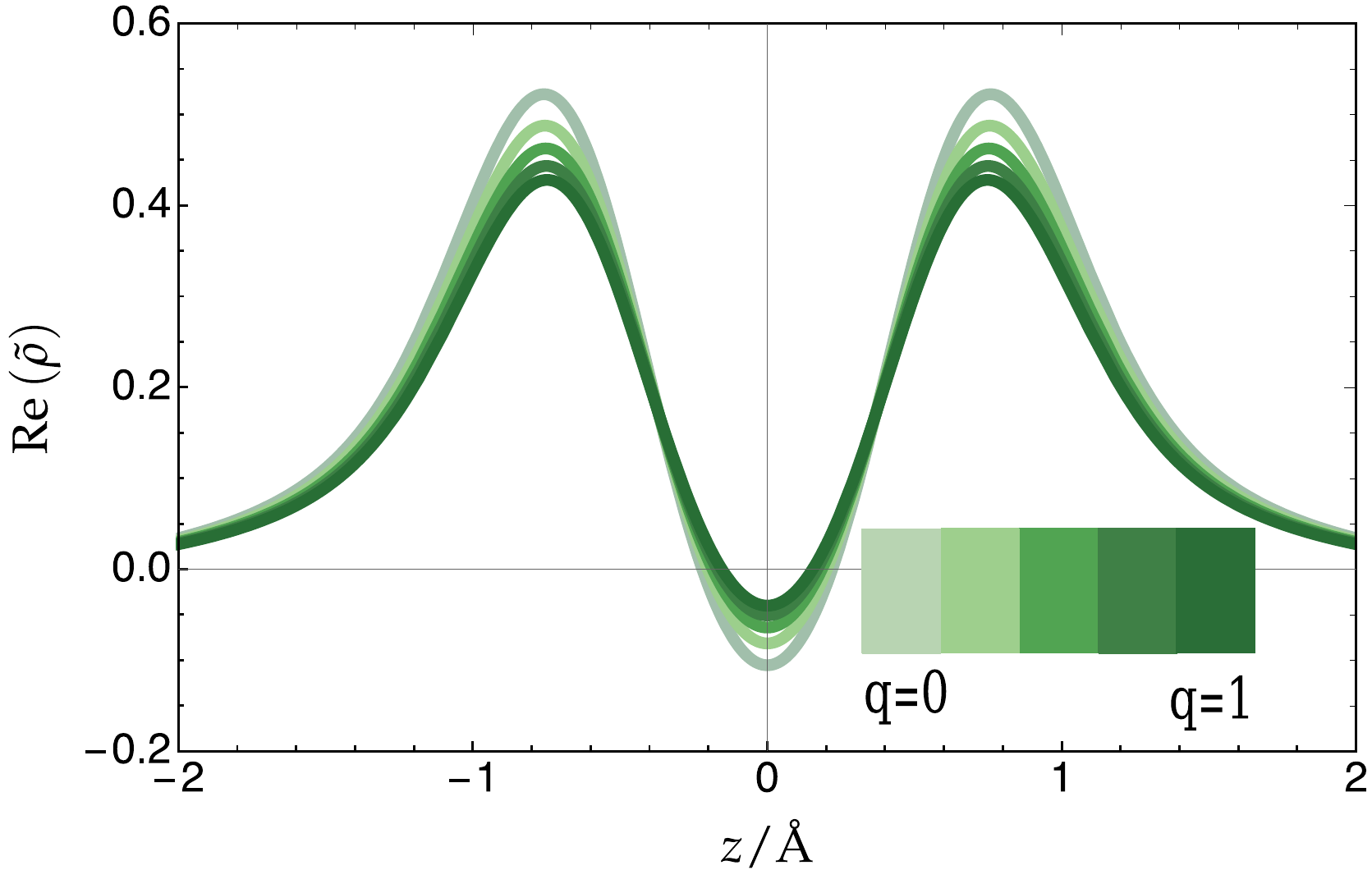}
		\subcaption{}    
		\label{subfig:hRKS_re_dens_R070}
	\end{subfigure}
	\begin{subfigure}[b]{.8\columnwidth}  
		\centering 
		\hspace{-3em}\includegraphics[width=\columnwidth]{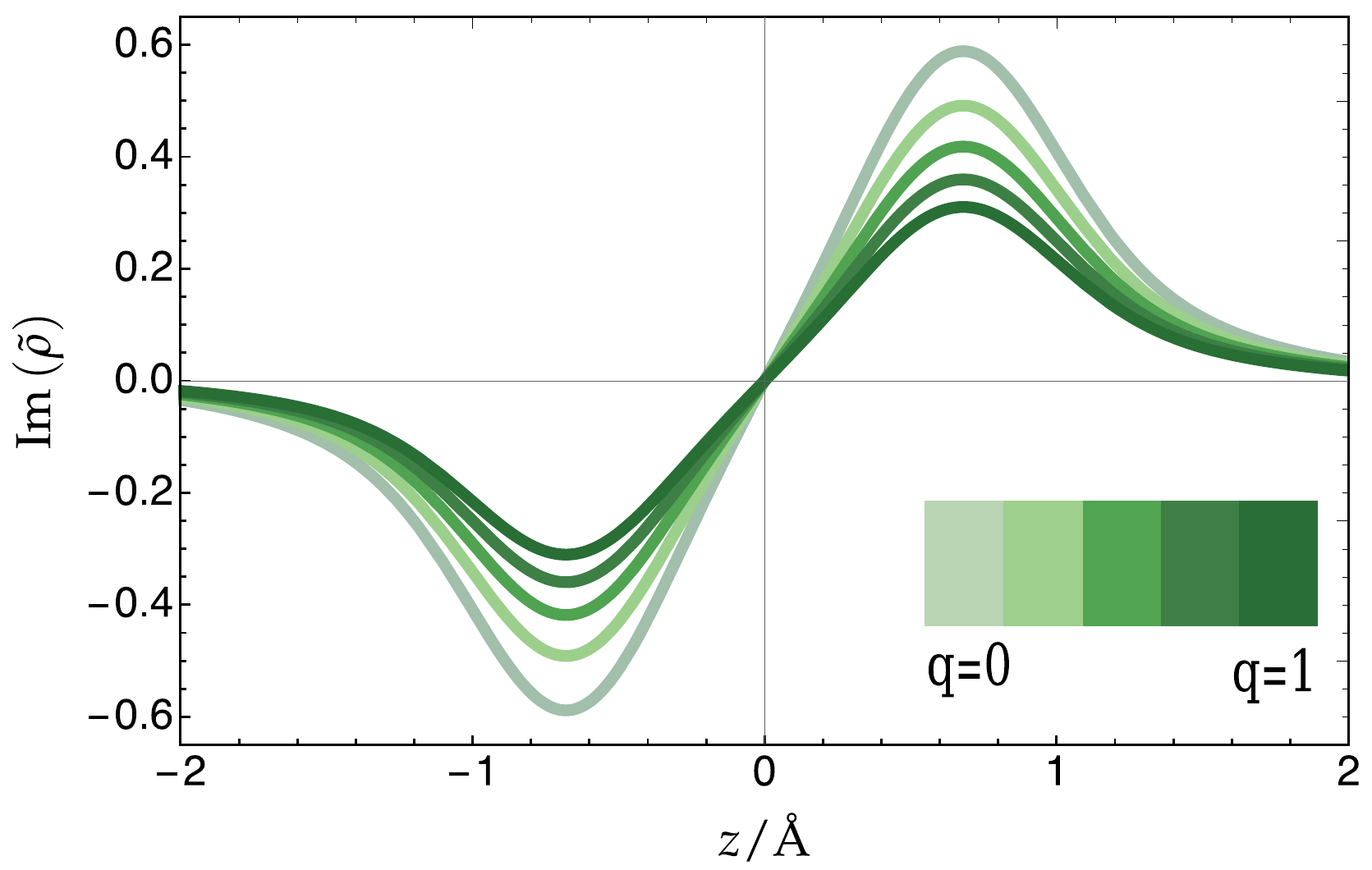}
		\subcaption{}
		\label{subfig:hRKS_im_dens_R070}
	\end{subfigure}
	\begin{subfigure}[b]{.8\columnwidth}
		\centering
		\hspace{-3em}\includegraphics[width=\columnwidth]{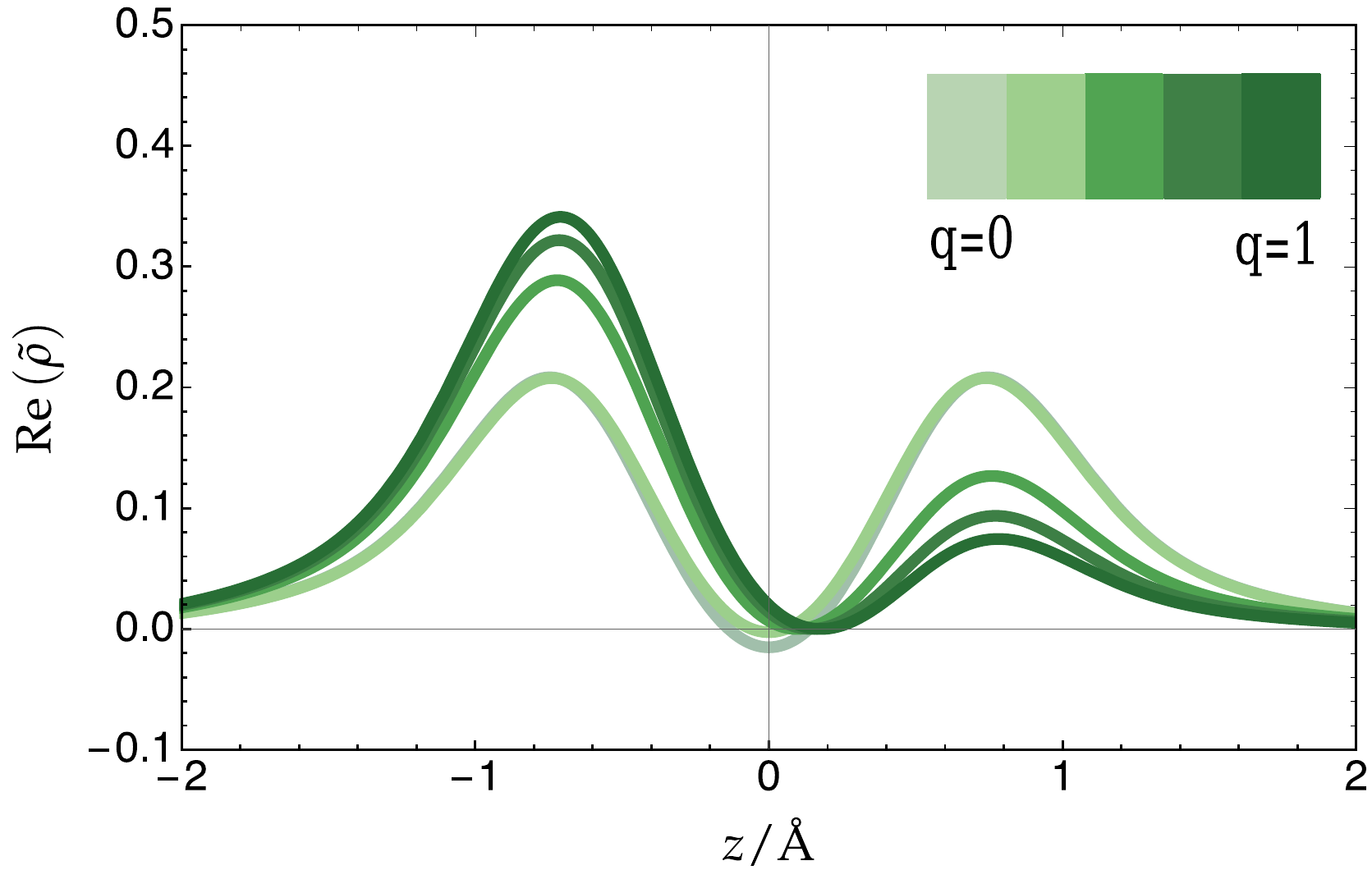}
		\subcaption{}    
		\label{subfig:hRKS_re_dens_R100}
	\end{subfigure}
	\begin{subfigure}[b]{.8\columnwidth}
		\centering
		\hspace{-3em}\includegraphics[width=\columnwidth]{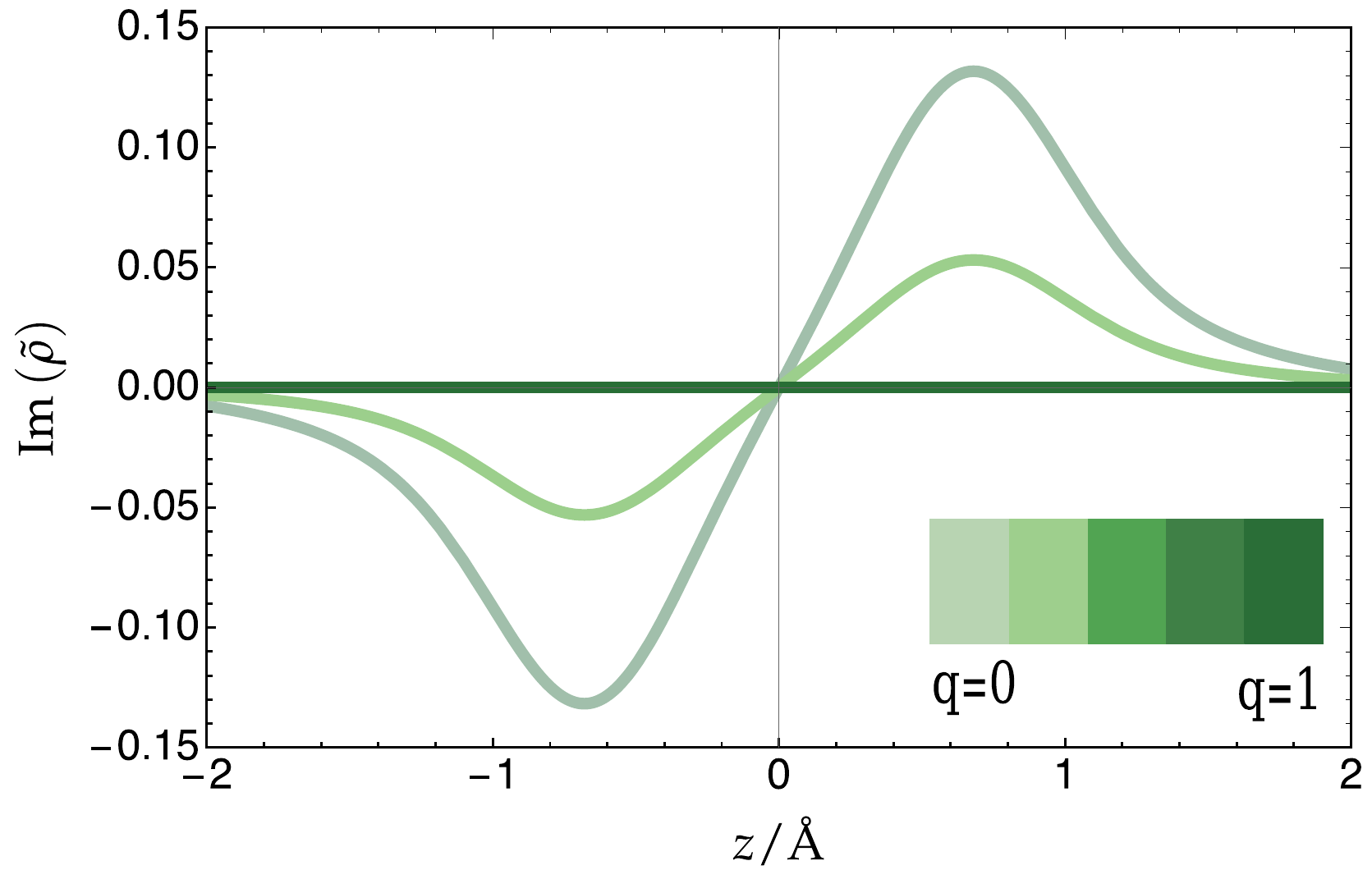}
		\subcaption{}    
		\label{subfig:hRKS_im_dens_R100}
	\end{subfigure}
	\caption{Real (\subref{subfig:hRKS_re_dens_R070}) and imaginary (\subref{subfig:hRKS_im_dens_R070}) components of the holomorphic electron density $\tilde{\rho}\qty(\br)$ for the h-RKS solution at different scaling factors $q$ (bond length: $\SI{0.70}{\angstrom}$). 
    Real (\subref{subfig:hRKS_re_dens_R100}) and imaginary (\subref{subfig:hRKS_im_dens_R100}) components of the holomorphic electron density $\tilde{\rho}\qty(\br)$ for the h-RKS solution at different scaling factors $q$ (bond length: $\SI{1.10}{\angstrom}$). 
    The internuclear axis is aligned with the $z$-axis.
	}
	\label{fig:hRKSdensityChange}
\end{figure*}

The iterative h-KS approach now allows the evolution of all h-HF solutions into the h-LDA functional to be directly visualised. 
The corresponding closed-shell density for the holomorphic ionic state is illustrated at bond lengths of $0.70$ and $\SI{1.10}{\angstrom}$ in
Figure~\ref{fig:hRKSdensityChange}.
At the intermediate bond length $\SI{1.10}{\angstrom}$ between the two coalescence points,  
there is now a clear evolution from the complex-valued h-HF density to the real-valued ionic KS-LDA density, confirming
that the two solutions are linked across the SCF approximation.
Comparing Figures~\ref{adiabat_change} and \ref{fig:hRKSdensityChange} demonstrates that the h-RKS solutions at both $0.70$ and $\SI{1.10}{\angstrom}$
show a more pronounced relaxation of the holomorphic density between HF and LDA than the density-fitting. 
This greater relaxation is to be expected if the density-fitting basis is not large enough to adequately fit the true holomorphic density, as described in Section~\ref{sec:DensityFitting}.

The rigorous map between multiple HF and DFT solutions finally allows us to understand how the change in exchange-correlation
potential affects the existence of symmetry-broken SCF solutions in \ce{H2}. 
In particular, the coalescence point for the unrestricted symmetry-broken solution  
occurs at a longer bond length in LDA than HF, in contrast to the coalescence for symmetry-broken ionic solution
that disappears at a shorter bond length in LDA than HF.
This observation suggests that the LDA exchange-correlation functional disfavours the spin-symmetry breaking of
the low-energy unrestricted solution, in line with our previous conclusions in the model electron transfer 
system (Section~\ref{sec:motivation}). 

\section{Concluding Remarks}
\label{conclusion}

The HF approximation and DFT are often considered as distinct methods in electronic structure. 
However, both approaches intimately linked  through the SCF approximation.
While it is well known that the self-consistency of HF theory can yield several optimal solutions, which may correspond to physically distinct electronic states, the 
existence of multiple DFT solutions is far less understood.
Furthermore, to the best our knowledge, direct connections between multiple HF and DFT solutions have never previously been explored.

In this work, we have performed a first investigation into the mapping of multiple SCF solutions between the HF and DFT energy surfaces.
Using a model electron transfer system,\cite{Jensen2018} we have found that the three low-lying HF states representing diabatic electron transfer configurations coalesce onto one DFT state.
This single DFT solution appears to be adiabatic in nature and maps continuously onto the lowest energy HF state at any geometry. 
As the SCF approximation is scaled between HF and DFT, we have shown that the disappearance of two SCF solutions states occurs in an analogous way to the coalescence of 
real HF states as the molecular structure changes.
Furthermore, we have shown that the coalescence of these SCF states is induced by an overall reduction in the exchange interaction between HF and the corresponding DFT functional,
highlighting the effect that this energy contribution has in driving spin-symmetry breaking.

To extend SCF solutions across all molecular structures and exchange-correlation functionals, we have developed two complex-analytic holomorphic extensions that can be applied in both 
the HF and DFT frameworks.
from the conventional electron density.
The first approach, based on a density-fitting approximation, allows the DFT energy to be expressed a complex-analytic polynomial functional and provides a mathematically 
rigorous extension to h-HF theory for guaranteeing the existence of multiple solutions.
However, solving the density-fitting equations relies on the introduction of auxiliary basis sets and appears to suffer from severe numerical challenges.
In contrast, the second approach considers a ``na\"{i}ve'' extension of h-HF theory whereby the KS-DFT equations are self-consistently solved using the complex-symmetric
holomorphic density. 
This h-KS approach requires minimal modifications to a standard KS-DFT implementation, and appears to allow complex-valued holomorphic DFT solutions to be uniquely identified
beyond the coalescence points where real DFT solutions coalesce and vanish.

Using the h-DFT method, we have investigated the complete mapping of the closed-shell states in \ce{H2} between HF theory and LDA-DFT.
By considering both restricted and unrestricted SCF solutions, we have identified fundamental differences in the way that different types of symmetry breaking evolve between HF theory and DFT.
In particular, spin-symmetry breaking in the ground-state unrestricted SCF solution appears to be discouraged using the LDA functional, with the coalescence point occurring at a larger bond length than in the HF approximation.
On the other hand, spatial-symmetry breaking in the ionic restricted SCF solution occurs at a shorter bond length in the LDA functional.
Alongside the model electron transfer model, these results suggest that different types of symmetry breaking are induced by changes in different relative components of the energy, such as the 
exchange interaction in the spin-symmetry-broken UHF solution. 

The nature of multiple DFT solutions, and their relationship to multiple HF states, is only beginning to be understood.
Like any single-determinant theory, one of the major deficiencies of DFT is the challenge of accurately describing static correlation effects.\cite{DFTdevelopmentReview}
Linear expansions of multiple DFT solutions have already been proposed as one extension beyond the single-reference DFT approximation,\cite{CDFT-CI} 
providing a direct analogy to the construction of multireference NOCI wave functions using multiple HF solutions.\cite{Thom2009}
Complex-valued h-HF solutions have been essential in the development of continuous NOCI basis sets across all molecular structures,\cite{Burton2019c} and it is 
likely that h-DFT solutions will serve a similar purpose for multireference DFT expansions.
Furthermore, understanding exactly how the choice of DFT functional affects the existence or coalescence of SCF solutions will be essential if multiple DFT solutions
are to be routinely used to interpret chemical processes, and we hope to continue this investigation in future publications.

Finally, it has been suggested that  any modern density-functional approximation must, to some extent, include nonlocality in the exchange description, 
as provided exactly by HF theory.\cite{DFTperspective}
While hybrid functionals achieve this nonlocality by empirically mixing local and nonlocal energy contributions, other density functionals employ 
use range separation or long-range correction. \cite{B3LYP, LR-DFT1, LR-DFT2, LR-DFT3, range_separated+MR, range_separated}
However, the choice of mixing or range is ultimately made by fitting to empirical data.
In contrast, we have revealed that the existence of symmetry-broken DFT solutions that include local or nonlocal electron densities is strongly dependent on the relative strengths 
of the Coulomb and exchange contributions, and is heavily influenced by the choice of exchange-crrelation functional.
We therefore believe that, by establishing a universal framework of multiple SCF states across all molecular structures or exchange-correlation functionals, the h-DFT approach will 
provide an entirely new approach for identifying the ``ideal'' combination of locality and nonlocality in DFT functionals through a theoretically justified, rather than empirical, foundation.

\section*{Acknowledgements}

HGAB acknowledges the Cambridge Trust for a PhD scholarship. AJWT thanks the Royal Society
for a University Research Fellowship (UF110161).

\section*{Supporting Information}
Numerical results for the h-RKS solutions of \ce{H2} using the minimal STO-3G basis set at bond lengths of $4.00$, $1.10$, and $\SI{0.70}{\angstrom}$.

\begin{appendix}
\section{Implementation of Holomorphic Density Fitting} 
\label{sec:appendix}
The holomorphic DFT as presented here relies on the method of density-fitting. This requires the introduction of an additional basis set for fitting the density, with basis functions $\xi_{\alpha}(\mathbf{r})$ and density-fitting coefficients $\{f^{\alpha}\}$.  
Due to the form of the LDA exchange functional, it is helpful to fit the cube-root of the density to retain sets of polynomial equations, rather than the density itself, giving
\begin{equation}
\tilde{\rho}\qty(\br)^{\frac13} = \sum_{\alpha}^{\infty} f^{\alpha} \xi_{\alpha}(\mathbf{r}).
\label{cubedRootDen}
\end{equation}
Following the form of the Kohn--Sham energy, the holomorphic Kohn--Sham energy is given by \begin{equation}
    \tilde{E}^{\text{KS}}[\tilde{\rho}]=\tilde{T}_{\text{s}}[\tilde{\rho}]+\tilde{E}_{\text{eN}}[\tilde{\rho}]+\tilde{E}_{\text{J}}[\tilde{\rho}]+\tilde{E}_{\text{XC}}[\tilde{\rho}],
\end{equation}
consisting of non-interacting kinetic, electron-nuclear, and electron-electron Coulomb and exchange-correlation terms respectively.  The holomorphic form of the latter three is identical to the conventional expressions, merely substituting the holomorphic density for the real density.

The non-interacting kinetic energy, $\tilde{T}_{\text{s}}$, however,
cannot be expressed polynomially in terms of the DF coefficients, because it depends explicitly upon MO coefficients, $\{c_{\cdot i}^{\mu \cdot}\}$ of the occupied Kohn--Sham orbitals, and only implicitly depends on the holomorphic electron density $\tilde{\rho}$.
In a two-electron system using a doubly occupied orbital, the non-interacting kinetic energy can be expressed exactly in terms of the density as $\frac{1}{8}\int \frac{|\mathbf{\nabla}\rho|^2}{\rho} d^3{\mathbf{r}}$, but this form is not amenable to a polynomial form.

We therefore express the holomorphic electronic energy as a function of both the MO coefficients $\{c_{\cdot i}^{\mu \cdot}\}$ and the set of density-fitting coefficients $\{f^{\alpha}\}$.
We constrain these coefficients by attempting to equate the holomorphic density derived from the density-fitting with that derived from the MO coefficients.  
This constraint defines a unique relation between the two expansion sets given by
\begin{align}
\sum_{\mu \nu}^n c_{\cdot 1}^{\mu\cdot} c_{1\cdot}^{\cdot\nu} \chi_{\mu}(\mathbf{r}) \chi_{\nu}(\mathbf{r}) 
&= 
\sum_{\alpha \beta \gamma}^m f^{\alpha} f^{\beta} f^{\gamma} \xi_{\alpha}(\mathbf{r}) \xi_{\beta}(\mathbf{r}) \xi_{\gamma}(\mathbf{r}) 
\label{ConnectBasises1}.
\end{align}
In addition, we equate the total derivative of the density with respect to one MO coefficient to derive the additional relation
\begin{equation}
\begin{split}
\frac{\text{d}}{\text{d} c_{\cdot 1}^{\mu\cdot}}
\sum_{\mu \nu}^n c_{\cdot 1}^{\mu\cdot} c_{1\cdot }^{\cdot \nu} \chi_{\mu}(\mathbf{r}) \chi_{\nu}(\mathbf{r}) &= 
\\
\frac{\text{d}}{\text{d} c_{\cdot 1}^{\mu\cdot}} \sum_{\alpha \beta \gamma}^m &f^{\alpha} f^{\beta} f^{\gamma} \xi_{\alpha}(\mathbf{r}) \xi_{\beta}(\mathbf{r}) \xi_{\gamma}(\mathbf{r}), 
\end{split}
\label{ConnectBasises2}
\end{equation}
that, together with Eq.~\eqref{ConnectBasises1}, fully determines the set of density-fitting coefficients.

To illustrate our approach, consider a two-electron system where the electronic energy is described in terms of both sets of coefficients as $\tilde{E}(\{c_{\cdot 1}^{\mu \cdot}\}, \{f^{\alpha}\})$.  
The lower index of $c$ is restricted to $1$ as there is only one doubly-occupied orbital. 
The cube-root density given by $\tilde{\rho}\qty(\br)^{\frac{1}{3}} = \sum_{\alpha}^m f^{\alpha} \xi_{\alpha}(\mathbf{r})$ and the occupied orbital is defined as $\phi(\mathbf{r}) = \sum^n_{\mu} c^{\mu\cdot}_{\cdot 1} \chi_{\mu}(\mathbf{r})$.
The polynomial system of equations required to describe the self-consistent field therefore contains $n+m$ unknowns, and thus identifying a self-consistent solution requires $n+m$ relations .

To set up the determined polynomial system, we derive the set of equations required for this two-electron, one-orbital system using two AO basis functions. 
The set of MO coefficients is given explicitly by $\{c_{\cdot i}^{\mu \cdot}\} = \{c_{\cdot 1}^{1 \cdot}, c_{\cdot 1}^{2 \cdot}\}$.  
The stationary condition if the holomorphic energy is then given by the vanishing derivatives,
\begin{align}
\frac{\text{d} E(\{c_{\cdot 1}^{\mu\cdot}\}, \{f^{\alpha}\})}{\text{d} c_{\cdot 1}^{\mu\cdot}} &= 0 \label{extremal},
\end{align}
and the normalization of the orbital leads to the constraint
\begin{align}
\sum_{\mu \nu}^n c_{1\cdot}^{\cdot\mu} S_{\mu\nu} c_{\cdot 1}^{\nu\cdot} &= 1
\label{ortho},
\end{align}
where $S_{\mu\nu}=\langle\chi_\mu|\chi_\nu\rangle$ is the AO overlap matrix.  
The implicit relationship between the two MO coefficients defined by Eq.~\eqref{ortho} allows stationary points to be identified with the
total derivative of the energy with respect to only one orbital coefficient, as given by Eq.~\eqref{extremal}.
Furthermore, the derivatives with respect to the density-fitting coefficients can be derived by exploiting the relationships \eqref{ConnectBasises1} and \eqref{ConnectBasises2}.

Due to the incompleteness of the density-fitting basis, it may not always be possible to satisfy Eq.~\eqref{ConnectBasises1}.
Therefore, to solve \eqref{ConnectBasises1} and \eqref{ConnectBasises2}, they are projected onto another auxiliary basis $\ket{\tau}$ defined as
\begin{equation}
\ket{\tau(\mathbf{r})} = \{\tau_1(\mathbf{r}), ... , \tau_l(\mathbf{r})\}
\end{equation}
where the size $l$ of the basis $\ket{\tau}$ is defined as $l = n + m - 2$ for a one-orbital two-electron system.

The projection of relations \eqref{ConnectBasises1} and \eqref{ConnectBasises2} onto the $\lambda^\text{th}$ basis function $\ket{\tau_{\lambda}}$ leads to
\begin{align}
\sum_{\mu \nu}^n c_{\cdot 1}^{\mu} c_{\cdot 1}^{\nu} \langle \tau_{\lambda} | \chi_{\mu} \chi_{\nu} \rangle 
&= \sum_{\alpha \beta \gamma}^m f^{\alpha} f^{\beta} f^{\gamma}\langle \tau_{\lambda} | \xi_{\alpha} \xi_{\beta}\xi_{\gamma} \rangle 
\\
\frac{\text{d}}{\text{d} c_{\cdot 1}^{\mu\cdot}}
\sum_{\mu \nu}^n c_{\cdot 1}^{\mu\cdot} c_{1\cdot}^{\cdot\nu} \langle \tau_{\lambda} | \chi_{\mu} \chi_{\nu} \rangle 
&=
\frac{\text{d}}{\text{d} c_{\cdot 1}^{\mu}} \sum_{\alpha \beta \gamma}^m f^{\alpha} f^{\beta} f^{\gamma} \langle \tau_{\lambda} | \xi_{\alpha} \xi_{\beta} \xi_{\gamma} \rangle.
\end{align}

Finally, the basis functions $\xi(\mathbf{r})$ of the density-fitting expansion and the auxiliary basis $\tau(\mathbf{r})$ are defined in relation to the atomic-orbital basis set $\chi(\mathbf{r})$.
For the investigated example of LDA-DFT, a possible choice of the density-fitting basis $\xi(\mathbf{r})$ in relation to the AO basis $\chi(\mathbf{r})$ is $\xi(\mathbf{r}) =  \chi^{2/3}(\mathbf{r})$, with a suitable normalisation factor to ensure $\langle \tau_i | \xi_i \xi_i \xi_i \rangle = 1$. 
The projection basis $\tau(\mathbf{r})$ can then be chosen such that $\tau(\mathbf{r}) = \chi^2(\mathbf{r})$.
\end{appendix}

\section*{References}
\bibliography{main}

\end{document}


\title{Holomorphic Density-Functional Theory --- Supporting Information}

\author{Rhiannon A.~Zarotiadis}
\email{rhiannon.zarotiadis@phys.chem.ethz.ch}
\affiliation{\UCAM}
\affiliation{\ETHZ}

\author{Hugh~G.~A.~Burton}
\email{hb407@cam.ac.uk}
\affiliation{\UCAM}

\author{Alex J.~W.~Thom}
\email{ajwt3@cam.ac.uk}
\affiliation{\UCAM}

\date{\today}

\maketitle

\vspace{-3em}
\section{Holomorphic Kohn--Sham Solutions of \ce{H2} (STO-3G)}

In this Section, we provide numerical data for the holomorphic Kohn--Sham (h-KS) solutions
identified in the minimal basis \ce{H2} molecule using the LDA exchange functional.\cite{bookLDA}
Molecular orbital coefficients of stationary states are represented in the atomic orbital STO-3G basis.
All stationary points are converged to a DIIS error of $10^{-8}$.
The holomorphic energy may be complex-valued in general, but in each case described, this energy remains purely real. 

\begin{table}[h!]
\begin{tabular}{ S[table-format=4.8] S[table-format=4.8] S[table-format=4.8] }
\hline\hline
{$C^{1 \cdot}_{\cdot 1}$} & {$C^{2 \cdot}_{\cdot 1}$} & {$\text{Re}[\text{Energy}] / \text{E}_{\text{h}}$} 
\\ \hline
0.706482 & 0.706482 & -0.688232
\\
0.707733 & -0.707733 & -0.683115
\\
-0.002607 & 1.000000 & -0.180309
\\
1.000000 & -0.002607 & -0.180309
\\
\hline \hline
\end{tabular}
\caption{Molecular orbital coefficients and holomorphic energies for the four h-RKS solutions
 at $R=\SI{4.00}{\angstrom}$. Every solution corresponds exactly with a real RKS stationary point.}
\end{table}

\begin{table}[h!]
\begin{tabular}{ S[table-format=4.8] S[table-format=4.8] S[table-format=4.8] }
\hline\hline
{$C^{1 \cdot}_{\cdot 1}$} & {$C^{2 \cdot}_{\cdot 1}$} & {$\text{Re}[\text{Energy}] / \text{E}_{\text{h}}$} 
\\ \hline
0.589341 & 0.589341 & -0.968592
\\
0.944561 & -0.944561 & -0.105412
\\
1.090490 & -0.680902 & -0.086347
\\
-0.680902 & 1.090490 & -0.086347
\\
\hline \hline
\end{tabular}
\caption{Molecular orbital coefficients and holomorphic energies for the four h-RKS solutions
 at $R=\SI{1.10}{\angstrom}$. Every solution corresponds exactly with a real RKS stationary point.}
\end{table}

\begin{table}[h!]
\begin{tabular}{ S[table-format=7.11] S[table-format=7.11] S[table-format=4.8] }
\hline\hline
{$C^{1 \cdot}_{\cdot 1}$} & {$C^{2 \cdot}_{\cdot 1}$} & {$\text{Re}[\text{Energy}] / \text{E}_{\text{h}}$} 
\\ \hline
0.544559 & 0.544559 & -1.023440
\\
1.262060 & -1.262060 & -0.587391
\\
{$1.431470+0.291466 \text{i}$} & {$-1.431470+0.291466 \text{i}$} & 0.678579
\\
{$1.431470-0.291466 \text{i}$} & {$-1.431470-0.291466 \text{i}$} & 0.678579
\\
\hline \hline
\end{tabular}
\caption{Molecular orbital coefficients and holomorphic energies for the four h-RKS solutions
 at $R=\SI{0.70}{\angstrom}$. 
 Two solutions have complex-valued orbital coefficients and occur in a complex-conjugate pair.
 The energy of these complex-valued solutions is strictly real as these stationary points satisfy
 parity-time symmetry.\cite{Burton2019b}
 }
\end{table}


\section*{References}
\bibliography{main}